\documentstyle[12pt]{article}
\newcommand{\nc}{\newcommand}
\nc{\al}{\alpha}
\nc{\g}{\gamma}
\nc{\G}{\Gamma}
\nc{\D}{\Delta}
\nc{\ald}{{\dot \al}}
\nc{\betad}{{\dot \beta}}
\nc{\gd}{{\dot \gamma}}
\nc{\sigmad}{{\dot \sigma}}
\nc{\mud}{{\dot \mu}}
\nc{\aldd}{{\ddot \al}}
\nc{\betadd}{{\ddot \beta}}
\nc{\gdd}{{\ddot \gamma}}
\nc{\sigmadd}{{\ddot \sigma}}
\nc{\mudd}{{\ddot \mu}}
\nc{\la}{\lambda}
\nc{\La}{\Lambda}
\nc{\var}{\varphi}
\nc{\ba}{\beta_\al}
\nc{\bb}{\beta_\beta}
\nc{\ga}{\g^\al}
\nc{\gb}{\g^\beta}
\nc{\kvt}{\sqrt{t}}
\nc{\hn}{h^\vee}
\nc{\kn}{k^\vee}
\nc{\pa}{\partial}
\nc{\nn}{\nonumber \\ }
\nc{\hf}{\frac{1}{2}}         
\nc{\paj}{P_{-\al}^j}
\nc{\vmab}{V_{-\al}^\beta}
\nc{\vab}{V_\al^\beta}
\nc{\vib}{V_i^\beta}
\nc{\db}{\pa_\beta}
\nc{\dtb}{\delta_\theta^\beta}
\nc{\binomial}[2]{\left (\begin{array}{c} {#1}\\ {#2} \end{array}
\right )}
\nc{\ben}{\begin{equation}}
\nc{\een}{\end{equation}}
\nc{\bea}{\begin{eqnarray}}
\nc{\eea}{\end{eqnarray}}
\nc{\bra}[1]{\langle {#1}|}
\nc{\ket}[1]{|{#1}\rangle}
\nc{\cpp}{{C_+}^+}
\nc{\cnp}{{C_0}^+}
\nc{\cmp}{{C_-}^+}
\nc{\cmn}{{C_-}^0}
\nc{\cmm}{{C_-}^-}
\nc{\vpp}{{V_+}^+}
\nc{\C}{\mbox{\hspace{1.24mm}\rule{0.2mm}{2.5mm}\hspace{-2.7mm} C}}
\nc{\Nat}{\mbox{\hspace{.04mm}\rule{0.2mm}{2.8mm}\hspace{-1.5mm} N}}
\nc{\spa}{\hspace{1 cm},\hspace{1 cm}}
\nc{\vs}{\vspace}
\nc{\NP}[1]{Nucl.\ Phys.\ {\bf #1}}
\nc{\PL}[1]{Phys.\ Lett.\ {\bf #1}}
\nc{\CMP}[1]{Commun.\ Math.\ Phys.\ {\bf #1}}
\nc{\PR}[1]{Phys.\ Rev.\ {\bf #1}}
\nc{\PRL}[1]{Phys.\ Rev.\ Lett.\ {\bf #1}}
\nc{\PTP}[1]{Prog.\ Theor.\ Phys.\ {\bf #1}}
\nc{\PTPS}[1]{Prog.\ Theor.\ Phys.\ Suppl.\ {\bf #1}}
\nc{\MPL}[1]{Mod.\ Phys.\ Lett.\ {\bf #1}}
\nc{\IJMP}[1]{Int.\ Jour.\ Mod.\ Phys.\ {\bf #1}}
\nc{\IM}[1]{Invent.\ Math.\ {\bf #1}}
\nc{\SJNP}[1]{Sov. J. Nucl. Phys.\ {\bf #1}}

\begin{document}

\topmargin -5mm
\oddsidemargin 5mm

\begin{titlepage}
\setcounter{page}{0}
\begin{flushright}
NBI-HE-97-15\\
June 1997
\end{flushright}

\vs{8mm}
\begin{center}
{\Large Free Field Realizations of Affine Current Superalgebras,}\\[.2cm]
{\Large Screening Currents and Primary Fields}

\vs{8mm}
{\large J{\o}rgen Rasmussen}\footnote{e-mail address: 
jrasmussen@nbi.dk}\\[.2cm]
{\em  The Niels Bohr Institute, Blegdamsvej 17, DK-2100 Copenhagen \O,
Denmark}\\[.5cm]

\end{center}

\vs{8mm}
\centerline{{\bf{Abstract}}}
In this paper free field realizations of affine current superalgebras are 
considered.
Based on quantizing differential operator realizations of the corresponding
basic Lie superalgebras, general and simple expressions for both the
bosonic and the fermionic currents are provided.
Screening currents of the first kind are also presented.
Finally, explicit free field realizations of primary fields with 
general, possibly non-integer, weights are worked out. A formalism is used
where the
(generally infinite) multiplet is replaced by a generating function primary
operator. The results allow setting up integral representations for 
correlators of primary fields corresponding to integrable representations.
The results are generalizations to superalgebras of a recent
work on free field realizations of affine current algebras by
Petersen, Yu and the present author.\\[.4cm]
{\em PACS:} 11.25.Hf\\
{\em Keywords:} Conformal field theory; Lie superalgebra; Affine current
superalgebra; Free field realizations

\end{titlepage}
\newpage
\renewcommand{\thefootnote}{\arabic{footnote}}
\setcounter{footnote}{0}

\section{Introduction}

Since the work by Wakimoto \cite{Wak} on free field realizations of affine
$SL(2)$ current algebra much effort has been made in obtaining similar 
constructions in the general case, a problem in principle solved by 
Feigin and Frenkel \cite{FF}. Recently two independent methods have led
to general and 
explicit solutions \cite{deBF,JR1,PRY3}. The method used by Petersen,
Yu and the present author \cite{JR1,PRY3} gives particularly simple
and compact free field realizations and is amenable of generalizing
to affine current 
superalgebras. Much less coherent results have been established 
so far in the case of
superalgebras. Free field realizations are only known in certain particular
cases (see e.g. \cite{BO2,Ito3,Ito2,BKT}). 
In this paper we present general and explicit free field
realizations of affine current superalgebras, generalizing the results in
\cite{JR1,PRY3}. The affine current superalgebras we consider
have basic Lie superalgebras as classical counterparts.

Free field realizations
enables one in principle to build integral representations for correlators in
conformal field theory \cite{DF,FZ,FGPP,An}. 
In a recent series of papers 
Petersen, Yu and the present author
have carried out such a study for conformal field theory based on
affine $SL(2)$ current algebra \cite{PRY1,PRY2}. It turns out that 
screening operators of both the first and the second
kinds are crucial for being able to treat the general case of degenerate 
representations \cite{KK} and in particular
admissible representations \cite{KW}. In that
connection it is also necessary to be able to handle fractional 
powers of free fields. Well defined rules for that have been
established also in \cite{PRY1,PRY2}. A particular
interest in these techniques is due to their close relationship with 
2D quantum gravity and string theory \cite{HY,AGSY}.

In order to generalize our work on affine $SL(2)$ current algebra to affine
current superalgebras (and eventually to be able to treat superstring theory
along the lines of \cite{HY,AGSY}), one needs not only free field realizations
of the affine currents but also of screening currents and primary fields.
In this paper we present such realizations of screening currents of the
{\em first} kind, only. However, that is expected to be sufficient for
treating {\em integrable} representations. Nevertheless, the free field
realizations we provide of primary fields are valid for general
representations.
For that purpose we use techniques based on ``super-triangular'' 
coordinates on representation spaces, which in turn makes our results
very compact.

The reduction of the results presented in this paper on Lie superalgebras
and affine current superalgebras to standard bosonic Lie algebras and
affine current algebras, is essentially by ``fermionic truncation''. By that
we mean disregarding all terms involving at least one of the odd
parameters: odd roots $\ald$, odd ``super-triangular'' coordinates 
$\theta^\ald$ or fermionic ghost fields $(b_\ald,c^\ald)$, see below.
Thus, the results presented in this paper are direct generalizations of
similar ones for purely bosonic algebras in \cite{JR1,PRY3}.

The paper is organized as follows. Section 2 serves to fix notation. Some
basic Lie superalgebra properties are reviewed and systems of free
bosonic and fermionic ghost fields and of free bosonic scalar fields are
discussed. We define our ``super-triangular'' coordinates.
A key object is the introduction of a matrix representation
of the raising part of the Lie superalgebra depending on those coordinates
in the adjoint representation, since the main new results in subsequent 
sections are expressed in terms of that matrix.

In Section 3 we work out explicitly certain Gauss decompositions leading to 
differential operator realizations of the Lie superalgebras. This is the
first main new result in this paper. We then discuss 
some polynomials later to become building blocks in construction of
screening currents in Section 5. 

In Section 4 the differential operator realization of a Lie superalgebra
derived in Section 3, 
is quantized to a (generalized) Wakimoto free field realization of the
corresponding affine current superalgebra. The non-trivial part consists in 
taking care of multiple contractions by adding anomalous terms to the
lowering operators. The full free field realization is the second
main new result in this paper.

In Section 5 free field realizations of screening currents of the first
kind are provided. This is the third main new result in this paper. The
set of screening operators depends on the choice of simple roots. In 
particular, the numbers of bosonic and fermionic screening currents
are equal to the numbers of even and odd simple roots, respectively.

In Section 6 we give a thorough discussion of primary fields using the
formalism based on ``super-triangular'' parameters. Simple and general
free field realizations of primary fields with arbitrary, possibly
non-integer, weights are derived. This is the fourth and final main
new result obtained in this paper.

In Section 7 we compare a subset of the results obtained in this paper with 
results known in the literature, by working out explicit examples.

Section 8 contains concluding remarks, while classical and quantum
polynomial identities following from the differential operator
realization and the free field realization, respectively, 
are listed in Appendix A. 

\section{Notation}
Let $\mbox{{\bf g}}^0$ and $\mbox{{\bf g}}^1$ denote the even and odd parts of
the basic Lie superalgebra $\mbox{{\bf g}}$ of rank $r$, see \cite{Kac1}
and references therein. 
$\D=\D^0\cup\D^1$ is the set of roots of 
$\mbox{{\bf g}}$ where $\D^0$ ($\D^1$) is the set of even (odd) roots.
A generic positive even (odd) root is written $\al\in\D_+^0$ ($\ald\in\D_+^1$),
and for such a root we write $\al>0$ ($\ald>0$). An arbitrary positive
root is written $\aldd\in\D_+$. A choice of a set of simple 
roots is written $\left\{\aldd_i\right\}_{i=1,...,r}$ in which 
there are $r^0$ even ones and $r^1=r-r^0$ odd ones.

Using the triangular decomposition 
\bea
 \mbox{{\bf g}}&=&\mbox{{\bf g}}_-\oplus\mbox{{\bf h}}\oplus\mbox{{\bf g}}_+\nn
 &=&(\mbox{{\bf g}}_-^1\oplus\mbox{{\bf g}}_-^0)\oplus\mbox{{\bf h}}\oplus
  (\mbox{{\bf g}}_+^0\oplus\mbox{{\bf g}}_+^1)
\eea
the raising and lowering even operators are denoted 
$E_\al\in$ {\bf g}$_+^0$ and $F_\al\in$ {\bf g}$_-^0$ respectively with 
$\al\in\Delta_+^0$, and 
$H_i\in$ {\bf h} are the Cartan operators. For the odd generators we use
lower case letters: $e_\ald\in$ {\bf g}$_+^1$ and $f_\ald\in$ {\bf g}$_-^1$,
where $\ald\in\D_+^1$. We let $J_a$ ($j_{{\dot a}}$) denote an even (odd)
generator. $J_A$ denotes an arbitrary Lie superalgebra generator. 
The (anti-)commutator algebra may be written
\ben
 {[}J_A,J_B{\}}={f_{A,B}}^CJ_C
\een
where ${[}\cdot,\cdot{\}}$ is an anti-commutator if both arguments are
fermionic, and otherwise a commutator. 
This is equivalent to
\ben
 {[}J_A,J_B{\}}=J_AJ_B-(-1)^{\mbox{{\footnotesize deg}}
  (J_A)\mbox{{\footnotesize deg}}(J_B)}J_BJ_A
\een
where the degree of a bosonic generator is deg$(J_a)=0$, while of a fermionic
generator it is deg$(j_{\dot{a}})=1$. Sometimes we will indicate the degree
by $p(A)=\mbox{deg}(J_A)$. Some of the structure coefficients 
satisfy
\bea
 {f_{\aldd_i,-\aldd_j}}^k&=&\delta_{ij}\delta_j^k\nn
 {f_{i,\pm\aldd}}^A&=&\pm\aldd(H_i)\delta_\aldd^A\nn
 {f_{\al,\beta}}^a&=&{f_{\al,\beta}}^{\al+\beta}\spa
  \mbox{for}\ \ \al+\beta\in\D^0\nn  
 {f_{\ald,\betad}}^a&=&{f_{\ald,\betad}}^{\ald+\betad}\spa
  \mbox{for}\ \ \ald+\betad\in\D^0\nn
 {f_{\al,\betad}}^{{\dot a}}&=&{f_{\al,\betad}}^{\al+\betad}\spa
  \mbox{for}\ \ \al+\betad\in\D^1
\label{Cartan-Weyl}
\eea 
while the Jacobi identity reads
\ben
 {[}J_A,{[}J_B,J_C{\}}{\}}={[}{[}J_A,J_B{\}},J_C{\}}
  +(-1)^{\mbox{{\footnotesize deg}}(J_A)
  \mbox{{\footnotesize deg}}(J_B)}{[}J_B,{[}J_A,J_C{\}}{\}}
\een
The Chevalley generators comprise the sets of raising and lowering operators
corresponding to simple roots and of Cartan generators, a total of $3r$ 
generators.
The Cartan-Killing form $\kappa$ satisfies
\bea
 \kappa_{AB}&=&(J_A,J_B)=(-1)^{\mbox{{\footnotesize deg}}
  (J_A)\mbox{{\footnotesize deg}}(J_B)}\kappa_{BA}\nn
 \kappa_{ij}&=&G_{ij}\spa\kappa_{\aldd,-\betadd}=0\ \ \mbox{unless}\ \ 
  \aldd=\betadd\nn
 2\hn\kappa_{AB}&=&str(\mbox{ad}_{J_A}\mbox{ad}_{J_B})=(-1)^{
  \mbox{{\footnotesize deg}}(J_B)}\left({f_{A,d}}^D{f_{B,D}}^d-
  {f_{A,\dot{d}}}^D{f_{B,D}}^{\dot{d}}\right)
\label{CKform}
\eea
where the metric $G_{ij}$ is related to the Cartan matrix 
$A_{ij}=\aldd_j(H_i)$ as $G_{ij}=A_{ij}\kappa_{\aldd_j,-\aldd_j}$.
$\hn$ is the dual Coxeter number of the Lie superalgebra.
We shall understand ``properly'' repeated indices as in
(\ref{CKform}) to be summed over. In the case of properly repeated
root indices ($\al,\beta,\g,...$)
the summation is over the positive (even and/or odd) roots.
In the explicit examples discussed in Section 7 we follow the normalization
convention of Kac \cite{Kac1}, saying that if $A_{ii}\neq0$ then $A_{ii}=2$,
and in the $i$'th row where $A_{ii}=0$ the first non-vanishing element of the
form $A_{i,i+j},\ j\geq1$ is 1. It is always possible to rescale the Lie 
superalgebra generators in order to meet these normalization conditions.
The bilinear form on the root space is defined by $\aldd_i\cdot\aldd_j=
(H_{\aldd_i},H_{\aldd_j})$, where $H_i=\kappa_{\aldd_i,-\aldd_i}H_{\aldd_i}$. 
The Weyl vector 
\bea
 &\rho=\rho^0-\rho^1&\nn
 \rho^0=\hf\sum_{\al>0}\al&,&\rho^1=\hf\sum_{\ald>0}\ald 
\eea
satisfies $\rho\cdot\aldd_i=\aldd_i^2/2$,
while the labels $\La_k$ and $\la_k$ of the weight $\La$ are defined by
\ben
 H_k\ket{\La}=\La(H_k)\ket{\La}=\La_k\ket{\La}
  \spa\La=\la_k\La^k\spa\La_i=\la_k\La^k(H_i)
\label{Dynkin}
\een
Here the particular (fundamental) set of linearly independent
weights $\{\La^k\}$ has the property
that the associated highest weight representations
$M_{\La^k}$ are finite dimensional.

Elements in $\mbox{\bf g}_+^0$ and $\mbox{\bf g}_+^1$ 
(or $\mbox{\bf g}_-$) or vectors in 
representation spaces (see below) are parametrized using ``super-triangular 
coordinates" denoted by $x^\al$ and (Grassmann odd variables)
$\theta^\ald$ respectively, 
one for each positive even or odd root. Thus we introduce the
Lie algebra elements
\ben
 g_+(x,\theta)=x^\al E_\al +\theta^\ald e_\ald\in \mbox{\bf g}_+ 
  \spa g_-(x,\theta)=x^\al F_\al +\theta^\ald f_\ald\in \mbox{\bf g}_-
\een
and the corresponding group elements $G_+(x,\theta)$ and $G_-(x,\theta)$:
\ben
 G_+(x,\theta)=e^{g_+(x,\theta)} \spa G_-(x,\theta)=e^{g_-(x,\theta)}
\een
The matrix representation $C(x,\theta)$ of $g_+(x,\theta)$ in the 
(pseudo-)adjoint representation is introduced as
\ben
 C_A^B(x,\theta)=-x^\beta {f_{\beta,A}}^B
  -\theta^\betad{f_{\betad,A}}^B
\label{cadj}
\een
This does not correspond precisely to the adjoint representation where
${(\mbox{ad}_{J_C})_A}^B={f_{A,C}}^B$ such that
\ben
 {(\mbox{ad}_{g_+(x,\theta)})_A}^B=-x^\beta{f_{\beta,A}}^B-(-1)^{p(A)}
  \theta^\betad{f_{\betad,A}}^B
\een
However, as it will be demonstrated
in subsequent sections, it is a very convenient matrix representation
essentially due to
\ben
 \left(\mbox{ad}_{g_+(x,\theta)}\right)^n(J_A)
  ={\left[(-C(x,\theta))^n\right]_A}^BJ_B
\een
The following notation is used for the (block) matrix elements
\ben
 C=\left(\begin{array}{lll}\cpp & 0 & 0\\
                 \cnp & 0 & 0\\
                 \cmp & \cmn & \cmm
         \end{array}  \right)
\label{C}
\een
${C_+}^+$ etc are matrices themselves. In ${C_+}^+$ both row and column 
indices are positive (even or odd) roots, in $\cmn$ the row index is a 
negative (even or odd) root and
the column index is a Cartan algebra index, etc.
One easily sees that (leaving out the arguments $x$ and $\theta$ 
for simplicity)
\bea
 {(C^n)_+}^+&=&({C_+}^+)^n\nn 
 {(C^n)_0}^+&=&{C_0}^+({C_+}^+)^{n-1}\nn
 {(C^n)_-}^0&=&({C_-}^-)^{n-1}{C_-}^0\nn
 {(C^n)_-}^-&=&({C_-}^-)^n\nn  
 {(C^n)_-}^+&=&\sum_{l=0}^{n-1}(\cmm)^l\cmp(\cpp)^{n-l-1}
  +\sum_{l=0}^{n-2}(\cmm)^l\cmn\cnp(\cpp)^{n-l-2}\nn
 0&=& {(C^n)_+}^0= {(C^n)_+}^-= {(C^n)_0}^0= {(C^n)_0}^- 
\eea
The block elements may be specified further as in
\ben
 {C_+}^+\sim\left(\begin{array}{ll}{C_{\ald}}^\betad & {C_{\ald}}^\beta \\
                 {C_{\al}}^\betad & {C_{\al}}^\beta 
                 \end{array}  \right)
 =\left(\begin{array}{ll}{-x^\g f_{\g,\ald}}^\betad & 
    {-\theta^{\gd}f_{\gd,\ald}}^\beta \\
                 {-\theta^{\gd}f_{\gd,\al}}^\betad & {-x^\g f_{\g,\al}}^\beta 
                 \end{array}  \right)
\een
We shall use repeatedly that $C_\al^\beta(x)$ vanishes unless 
$\al <\beta$, and similarly for the remaining 3 block elements of ${C_+}^+$.
This corresponds to each block (${C_{\ald}}^\betad$, ${C_{\ald}}^\beta$, 
${C_{\al}}^\betad$ and ${C_{\al}}^\beta$) being upper triangular with 
zeros in the diagonals. Likewise, the block elements of ${C_-}^-$ are 
lower triangular. 
It will turn out that we shall be able to provide remarkably simple universal
analytic expressions for most of the objects we consider, using the 
matrix $C(x,\theta)$.

For the associated affine Lie superalgebra, the operator product 
expansion, OPE, of the associated currents is
\ben
 J_A(z)J_B(w)=\frac{\kappa_{AB}k}{(z-w)^2}+\frac{{f_{A,B}}^C J_C(w)}{z-w}
\label{JAJB}
\een
where regular terms have been omitted. $k$ is the central extension or
level of the affine current superalgebra. We use the same notation
$J,E,F,H,j,e,f$ for the currents as for the algebra generators. Hopefully,
it will not lead to misunderstandings. 
The associated Sugawara construction is 
\ben
 T(z)=\frac{1}{2(k+\hn)}\kappa^{AB}:J_A(z)J_B(z):
\een
and has central charge
\ben
 c=\frac{k\ \mbox{sdim}(\mbox{{\bf g}})}{k+\hn}
\label{c}
\een
In the mode expansion
\ben
 J_A(z)=\sum_{n=-\infty}^\infty \hat{J}_{A,n}z^{-n-1}
\een
we use the identification
\ben
 \hat{J}_{A,0}\equiv J_A\in\ \mbox{{\bf g}} 
\label{identif}
\een

The standard free field construction 
(see \cite{Wak,FF,deBF,JR1,PRY3} for affine current algebras and 
\cite{BO2,Ito2} for affine current superalgebras)
consists in introducing for every positive {\em even}
root $\al>0$, a pair of free {\em bosonic} ghost
fields ($\ba,\ga$) of conformal weights (1,0) satisfying the OPE
\ben
 \ba(z)\gb(w)=\frac{{\delta_{\al}}^\beta}{z-w}
\een
The corresponding energy-momentum tensor is
\ben
 T_{\beta\g}=:\pa\ga\ba:
\label{Tbg}
\een
with central charge
\ben
 c_{\beta\g}=2|\D_+^0|=\mbox{dim}(\mbox{{\bf g}}^0)-r
\een
For every positive {\em odd} root $\ald>0$ one introduces a pair of free
{\em fermionic} ghost fields ($b_\ald,c^\ald$) of conformal weights
(1,0) satisfying the OPE
\ben
 b_\ald(z)c^\betad(w)=\frac{{\delta_{\ald}}^\betad}{z-w}
\een
The corresponding energy-momentum tensor is
\ben
 T_{bc}=:\pa c^\ald b_\ald:
\label{Tbc}
\een
with central charge
\ben
 c_{bc}=-2|\D_+^1|=-\mbox{dim}(\mbox{{\bf g}}^1)
\een
For every Cartan index $i=1,...,r$ one introduces a free scalar boson $\var_i$
with contraction
\ben
 \var_i(z)\var_j(w)=G_{ij}\ln(z-w)
\een
The energy-momentum tensor 
\ben
 T_\var=\hf:\pa\var\cdot\pa\var:-\frac{1}{\kvt}\rho\cdot\pa^2\var
\een
has central charge
\ben
 c_\var=r-\frac{\hn \mbox{sdim}(\mbox{{\bf g}})}{k+\hn}
\label{cvar}
\een
where the super-dimension sdim$(\mbox{{\bf g}})$ of the Lie superalgebra 
{\bf g} is defined as the difference dim({\bf g}$^0$) $-$ dim({\bf g}$^1$).
In obtaining (\ref{cvar}) we have used Freudenthal-de Vries (super-)strange
formula \cite{Kac2}
\ben
 \rho^2=\frac{\hn}{12}\mbox{sdim}(\mbox{{\bf g}})
\een
The total free field realization of the Sugawara energy-momentum tensor 
is $T=T_{\beta\g}+T_{bc}+T_\var$ and has indeed central charge (\ref{c}).
In subsequent sections, the combination $k+\hn$ will be abbreviated by $t$:
\ben
 t=k+\hn
\een
One of the new results in this paper will be explicit free field realizations 
of the currents in a general affine Lie superalgebra in Section 4, 
based on matrix representations similar to $C$ in (\ref{cadj}).

The vertex operator
\ben
 V_\La(z)=:e^{\frac{1}{\kvt}\La\cdot\var(z)}: 
\een
has conformal weight
\ben
 \D(V_\La)=\frac{1}{2t}(\La,\La+2\rho)
\een
It is also affine primary corresponding to highest weight $\La$. One of the
main new results in this paper will be an explicit general construction
of the {\em full} multiplet of primary fields, parametrized by the $x^\al$ and
$\theta^\ald$ coordinates in Section 6.

\section{Differential Operator Realization}

In this section we discuss differential operator realizations of the Lie
superalgebra {\bf g} on the polynomial ring $\C[x^\al,\theta^\ald]$. The
techniques for obtaining such realizations in the case of standard bosonic
Lie algebras have been known for some time (see e.g. \cite{Kos}) and
completed in \cite{JR1,PRY3}. The procedure and results of \cite{JR1,PRY3}
is generalized to cover Lie superalgebras in the following. In \cite{Ito2}
Ito used similar techniques to obtain differential operator realizations
of the {\em Chevalley generators}, see also Section 7.

The lowest weight vector $\bra{\La}$
in the (dual) representation space is introduced as 
\ben
 \bra{\La}F_\al=\bra{\La}f_\ald=0\spa\bra{\La}H_i=\La_i\bra{\La}
\een
An arbitrary vector in this representation space is parametrized as
\ben
 \bra{\La,x,\theta}=\bra{\La}G_+(x,\theta)
\een
The differential operator realization $\{\tilde{J}_A(x,\theta,\pa,\La)\}$
with $\pa_\al=\pa_{x^\al}$ and $\pa_\ald=\pa_{\theta^\ald}$ denoting partial
derivatives wrt $x^\al$ and $\theta^\ald$, is then defined by
\ben
 \bra{\La,x,\theta}J_A=\tilde{J}_A(x,\theta,\pa,\La)\bra{\La,x,\theta}
\een
and it follows immediately that the generators $\tilde{J}_A(x,\theta,\pa,\La)$
satisfy the 
Lie superalgebra commutation relations. It is convenient (in particular
when considering primary fields in Section 6) to have a similar 
notation for highest weight vectors
\bea
 \ket{\La,x,\theta}&=&G_-(x,\theta)\ket{\La}\nn
 J_A\ket{\La,x,\theta}&=&-J_A(x,\theta,\pa,\La)\ket{\La,x,\theta}
\label{ketdef}
\eea
The relation between the two sets of realizations of the Lie
superalgebra, $\{ \tilde{J}_A(x,\theta,\pa,\La)\}$ and
$\{ J_A(x,\theta,\pa,\La)\}$, is 
\ben
 \tilde{J}_A(x,\theta,\pa,\La)=-J_A^\dagger(x,\theta,\pa,\La)
\label{tilde}
\een
where the super-adjoint operation in {\bf g} is defined by
\bea
 E_\al^\dagger&=&F_\al\spa F_\al^\dagger=E_\al\spa H_i^\dagger=H_i\nn
 e_\ald^\dagger&=&f_\ald\spa f_\ald^\dagger=-e_\ald
\label{superadjoint}
\eea
In general \cite{NRS}, a super-adjoint (or grade adjoint)
operation is linear, degree (or grade) preserving and satisfies
\bea
 {[}J_A,J_B{\}}^\dagger&=&(-1)^{\mbox{{\footnotesize deg}}(J_A)
  \mbox{{\footnotesize deg}}(J_B)}{[}
  J_B^\dagger,J_A^\dagger{\}}\nn
 (J_A^\dagger)^\dagger&=&(-1)^{\mbox{{\footnotesize deg}}(J_A)}J_A
\eea
In particular, we have $(g_+(x,\theta))^\dagger=g_-(x,\theta)$ since
$(\theta^\ald J_A)^\dagger=\theta^\ald J_A^\dagger$.
Furthermore, one easily derives the following symmetries of the structure
coefficients
\bea
 {f_{-\al,\beta}}^{\pm\g}=-{f_{\al,-\beta}}^{\mp\g}\spa
  {f_{-\al,\betad}}^{\pm\gd}&=&\mp{f_{\al,-\betad}}^{\mp\gd}\spa
  {f_{-\ald,\betad}}^{\pm\g}={f_{\ald,-\betad}}^{\mp\g}\nn
 {f_{-\aldd,-\betadd}}^{-\gdd}&=&-{f_{\aldd,\betadd}}^\gdd
 \label{fsymm2}
\eea
Here all roots are meant to be positive.

The Gauss decompositions of $\bra{\La}G_+(x,\theta)e^{tJ_a}$ 
for $t$ small and of $\bra{\La}G_+(x,\theta)e^{\mu j_{\dot{a}}}$ for $\mu$ 
a Grassmann odd parameter, may be written
\bea
 \bra{\La}G_+(x,\theta)\exp(tE_\al)&=&\bra{\La}\exp\left( 
  g_+(x,\theta)+t\vab(x,\theta)E_\beta+tV_\al^\betad(x,\theta)e_\betad
  +{\cal O}(t^2)\right)\nn
 &=&\bra{\La}\exp\left(t\left(\vab(x,\theta)\db
  +V_\al^\betad(x,\theta)\pa_\betad\right)
  +{\cal O}(t^2)\right)G_+(x,\theta)\nn
 \bra{\La}G_+(x,\theta)\exp(tH_i)&=&\bra{\La}\exp\left(tH_i+{\cal O}(t^2)
  \right)\nn
 &\cdot&\exp\left(g_+(x,\theta)
  +tV_i^\beta(x,\theta)E_\beta+tV_i^\betad(x,\theta)
  e_\betad+{\cal O}(t^2)\right)\nn
 &=&\bra{\La}\exp\left(t\left(V_i^\beta(x,\theta)\db+V_i^\betad(x,\theta)
  \pa_\betad+\La_i\right)+{\cal O}(t^2)\right)\nn
 &\cdot&G_+(x,\theta)\nn 
 \bra{\La}G_+(x,\theta)\exp(tF_\al)&=&\bra{\La}\exp\left(
  tQ_{-\al}^{-\beta}(x,\theta)F_\beta+tQ_{-\al}^{-\betad}(x,\theta)f_\betad
  +{\cal O}(t^2)\right)\nn
 &\cdot&\exp\left(t\paj(x,\theta)H_j+{\cal O}(t^2)\right)\nn
 &\cdot&\exp\left(g_+(x,\theta)+t\vmab(x,\theta)E_\beta
  +tV_{-\al}^\betad(x,\theta)e_\betad+{\cal O}(t^2)\right)\nn
 &=&\bra{\La}
  \exp\left(t\left(\paj(x,\theta)\La_j\right.\right.\nn
 &+&\left.\left.\vmab(x,\theta)\db
  +V_{-\al}^\betad(x,\theta)\pa_\betad\right)
  +{\cal O}(t^2)\right)G_+(x,\theta)\nn
 \bra{\La}G_+(x,\theta)\exp(\mu e_\ald)&=&\bra{\La}\exp\left( 
  g_+(x,\theta)+\mu V_\ald^\beta(x,\theta)E_\beta+\mu 
  V_\ald^\betad(x,\theta)e_\betad\right)\nn
 &=&\bra{\La}\exp\left(\mu \left(V_\ald^\beta(x,\theta)\db
  +V_\ald^\betad(x,\theta)\pa_\betad\right)\right)G_+(x,\theta)\nn
 \bra{\La}G_+(x,\theta)\exp(\mu f_\ald)&=&\bra{\La}\exp\left(
  \mu Q_{-\ald}^{-\beta}(x,\theta)F_\beta+\mu 
  Q_{-\ald}^{-\betad}(x,\theta)f_\betad\right)\nn
 &\cdot&\exp\left(\mu P_{-\ald}^j(x,\theta)H_j\right)\nn
 &\cdot&\exp\left(g_+(x,\theta)+\mu V_{-\ald}^\beta(x,\theta)E_\beta
  +\mu V_{-\ald}^\betad(x,\theta)e_\betad\right)\nn
 &=&\bra{\La}
  \exp\left(\mu \left(P_{-\ald}^j(x,\theta)\La_j\right.\right.\nn
 &+&\left.\left.V_{-\ald}^\beta(x,\theta)\db
  +V_{-\ald}^\betad(x,\theta)\pa_\betad\right)\right)G_+(x,\theta)
\label{Gauss1}
\eea
It follows that the differential operator realization is of the form
\bea
 \tilde{E}_\al(x,\theta,\pa)&=&V_\al^\betadd(x,\theta)\pa_\betadd\nn
 \tilde{H}_i(x,\theta,\pa,\La)&=&V_i^\betadd(x,\theta)
  \pa_\betadd+\La_i\nn
 \tilde{F}_\al(x,\theta,\pa,\La)&=&
  V_{-\al}^\betadd(x,\theta)\pa_\betadd+
  \paj(x,\theta)\La_j\nn
 \tilde{e}_\ald(x,\theta,\pa)&=&V_\ald^\betadd(x,\theta)\pa_\betadd\nn
 \tilde{f}_\ald(x,\theta,\pa,\La)&=&
  V_{-\ald}^\betadd(x,\theta)\pa_\betadd
  +P_{-\ald}^j(x,\theta)\La_j
\label{defVP}
\eea
Since $\tilde{E}_\al(x,\theta,\pa,\La)=\tilde{E}_\al(x,\theta,\pa)$ and
$\tilde{e}_\ald(x,\theta,\pa,\La)=\tilde{e}_\ald(x,\theta,\pa)$ are 
independent of $\La$ they may be defined through a Gauss decomposition alone.

We shall now work out explicitly the relevant Gauss decompositions
in order to determine the polynomials $V$, $P$ and $Q$. For that purpose
we use repeatedly the
Campbell-Baker-Hausdorff formula (see e.g. \cite{JR1} for a proof)
\bea
 e^Ae^{tB}&=&\exp\left\{A+t\sum_{n\geq0}\frac{B_n}{n!}(-\mbox{ad}_A)^nB
  +{\cal O}(t^2)\right\}\nn
 e^Ae^{\mu b}&=&\exp\left\{A+\sum_{n\geq0}\frac{B_n}{n!}(-\mbox{ad}_A)^n
  (\mu b)\right\}
\eea
where $\mu$ is a Grassmann odd parameter and $b$ is an odd operator, and
where $B_n$ are the Bernoulli numbers with generating function $B(u)$ 
\bea
 B(u)&=&\frac{u}{e^u-1}=\sum_{n\geq0}\frac{B_n}{n!}u^n\nn
 (B(u))^{-1}&=&\frac{e^u-1}{u}=\sum_{n\geq0}\frac{1}{(n+1)!}u^n
\label{Ber}
\eea
It turns out that only slight modifications are needed of the techniques
employed in the recent work \cite{JR1,PRY3} by Petersen, Yu and the present
author on purely bosonic free field realizations of affine current algebras.
Utilizing the results of that work we find
\bea
  V_\aldd^\betadd(x,\theta)&=&\left[B(C(x,\theta))\right]_\aldd^\betadd\nn
  V_i^\betadd(x,\theta)&=&-\left[C(x,\theta)\right]_i^\betadd\nn
  V_{-\aldd}^\betadd(x,\theta)&=&\left[e^{-C(x,\theta)}\right]_{-\aldd}^\gdd
   \left[B(-C(x,\theta))\right]_\gdd^\betadd\nn
  P_{-\aldd}^j(x,\theta)&=&\left[e^{-C(x,\theta)}\right]_{-\aldd}^j\nn
  Q_{-\aldd}^{-\betadd}(x,\theta)&=&
   \left[e^{-C(x,\theta)}\right]_{-\aldd}^{-\betadd}
\label{pol}
\eea
Note that the expressions are valid for {\em all} positive roots and that
the summation ($\sum_{\gdd>0}$)
in $V_{-\aldd}^\betadd(x,\theta)$ is over {\em both} even
and odd (positive) roots.
Due to the fact that for any given Lie superalgebra the matrix $C(x,\theta)$ is
nilpotent, the formal power series in (\ref{pol})
all truncate and become polynomials.
We refer to \cite{JR1,PRY3} for further details. The explicit
polynomial expressions in (\ref{pol}) comprise the first main new result
in this paper, since they provide us with explicit differential
operator realizations (\ref{defVP}) and (\ref{tilde}) of the Lie superalgebra 
in question.

A possible generalization of the work \cite{deBF} by de Boer and Feh\'er
on affine current algebras, may be based on the following (super-)trace
\ben
 V_\aldd^\betadd(x,\theta)str\left(G_+^{-1}(x,\theta)\pa_\betadd G_+(x,\theta)
  F_\gdd\right)=2\hn\kappa_{\aldd,-\gdd}
\een
Here we have used the common notation $F_\gdd$ for the 
lowering generators corresponding to the positive root $\gdd$.
The expression follows immediately from the realizations 
(\ref{Gauss1}) and (\ref{defVP}) of 
$\tilde{E}_\al(x,\theta)$ and $\tilde{e}_\ald(x,\theta)$.
In \cite{deBF} essentially the purely bosonic counterpart of this trace
is introduced as a key object in their explicit Wakimoto construction.

\subsection{Differential Screening Operators}

We shall also be interested in screening currents in Section 5. They
are built from certain differential operators $S_\aldd$ 
to be defined presently. In the case of purely bosonic Lie algebras
similar operators are well known \cite{FF,BMP,Ito,ATY,deBF,JR1,PRY3}.

Let $t$ and $\mu$ be Grassmann even and odd parameters, respectively.
The operators 
\ben
 S_\aldd(x,\theta,\pa)=S_\aldd^\betadd(x,\theta)\pa_\betadd
\een
are then defined through the Gauss decompositions
\ben
 e^{tS_\al(x,\theta,\pa)}G_+(x,\theta)=e^{-tE_\al}G_+(x,\theta) \spa
 e^{\mu S_\ald(x,\theta,\pa)}G_+(x,\theta)=e^{-\mu e_\ald}G_+(x,\theta)
\een
and we find
\ben
 S_\aldd^\betadd(x,\theta)=-\left[B(-C(x,\theta))\right]_\aldd^\betadd
\een
It follows from associativity of the Lie supergroup, that these polynomials
satisfy the commutator relations
\bea 
 \left[\tilde{E}_\al(x,\theta,\pa),S_\betadd(x,\theta,\pa)\right]&=&0\nn
 \left[\tilde{H}_i(x,\theta,\pa),S_\betadd(x,\theta,\pa)\right]&=&
  \betadd(H_i)S_\betadd(x,\theta,\pa)\nn
 \left[\tilde{F}_\al(x,\theta,\pa),S_\beta(x,\theta,\pa)\right]
  &=&\beta(H_j)P_{-\al}^j(x,\theta)S_\beta(x,\theta,\pa)+
  Q_{-\al}^{-\g}(x,\theta){f_{\beta,-\g}}^j\La_j\nn
 &-&{f_{\beta,-\g}}^\sigma Q_{-\al}^{-\g}(x,\theta)S_\sigma(x,\theta,\pa)
  -{f_{\beta,-\gd}}^\sigmad Q_{-\al}^{-\gd}(x,\theta)S_\sigmad(x,\theta,\pa)\nn
 \left[\tilde{F}_\al(x,\theta,\pa),S_\betad(x,\theta,\pa)\right]
  &=&\betad(H_j)P_{-\al}^j(x,\theta)S_\betad(x,\theta,\pa)-
  Q_{-\al}^{-\gd}(x,\theta){f_{\betad,-\gd}}^j\La_j\nn
 &+&{f_{\betad,-\gd}}^\sigma Q_{-\al}^{-\gd}(x,\theta)S_\sigma(x,\theta,\pa)
  -{f_{\betad,-\g}}^\sigmad Q_{-\al}^{-\g}(x,\theta)S_\sigmad(x,\theta,\pa)\nn
 \left[\tilde{e}_\ald(x,\theta,\pa),S_\beta(x,\theta,\pa)\right]&=&0\nn
 \left\{\tilde{e}_\ald(x,\theta,\pa),S_\betad(x,\theta,\pa)\right\}&=&0\nn
 \left[\tilde{f}_\ald(x,\theta,\pa),S_\beta(x,\theta,\pa)\right]&=&
  \beta(H_j)P_{-\ald}^j(x,\theta)S_\beta(x,\theta,\pa)+
  Q_{-\ald}^{-\g}(x,\theta){f_{\beta,-\g}}^j\La_j\nn
 &-&{f_{\beta,-\g}}^\sigma Q_{-\ald}^{-\g}(x,\theta)S_\sigma(x,\theta,\pa)
  -{f_{\beta,-\gd}}^\sigmad Q_{-\ald}^{-\gd}(x,\theta) 
  S_\sigmad(x,\theta,\pa)\nn
 \left\{\tilde{f}_\ald(x,\theta,\pa),S_\betad(x,\theta,\pa)\right\}&=&
  \betad(H_j)P_{-\ald}^j(x,\theta)S_\betad(x,\theta,\pa)
  -Q_{-\ald}^{-\gd}(x,\theta){f_{\betad,-\gd}}^j\La_j\nn
 &+&{f_{\betad,-\gd}}^\sigma Q_{-\ald}^{-\gd}(x,\theta)S_\sigma(x,\theta,\pa)
  -{f_{\betad,-\g}}^\sigmad Q_{-\ald}^{-\g}(x,\theta)S_\sigmad(x,\theta,\pa)
  \nn
 \left[S_\al(x,\theta,\pa),S_\beta(x,\theta,\pa)\right]&=&
  {f_{\al,\beta}}^\g S_\g(x,\theta,\pa)\nn
 \left[S_\al(x,\theta,\pa),S_\betad(x,\theta,\pa)\right]&=&
  {f_{\al,\betad}}^\gd S_\gd(x,\theta,\pa)\nn
 \left\{S_\ald(x,\theta,\pa),S_\betad(x,\theta,\pa)\right\}&=&
  {f_{\ald,\betad}}^\g S_\g(x,\theta,\pa)
\label{Scomm}
\eea

Several non-trivial {\em classical} polynomial identities (as opposed to
{\em quantum} polynomial identities, see Section 4 and Appendix A)
may be derived from the
commutator relations using the differential operator realizations 
of the Lie superalgebra generators (\ref{defVP}) and (\ref{Scomm}).
The identities will be used in subsequent sections, and are listed in 
Appendix A.

\section{Wakimoto Free Field Realization}

In the case of purely bosonic affine current algebras, it is well known
how to obtain free field realizations (based on bosonic ghost pairs
($\beta_\al,\g^\al$) and bosonic scalars $\var_i$) from the ``fermionic
truncation'' of the differential
operator realization $\{\tilde{J}_a\}$, simply by the substitutions 
\cite{FF,GMMOS,BMP,Ito1,Ito,Ku,ATY,JR1,PRY3}
\ben
 \pa_\al\rightarrow\beta_\al(z)\spa x^\al\rightarrow\g^\al(z)
  \spa\La_i\rightarrow\kvt\pa\varphi_i(z)
\een
followed by adding anomalous terms $\pa\g^\beta(z)F_{\al\beta}(\g(z))$
to the lowering part. The natural generalization of that to the case
of affine current superalgebras is to make the substitutions
(see \cite{BO2} for the case of $OSp(1|2)$, and \cite{Ito2} for Chevalley
generators in general)
\bea
 \pa_\al\rightarrow\beta_\al(z)&\spa&x^\al\rightarrow\g^\al(z)
  \spa\La_i\rightarrow\kvt\pa\varphi_i(z)\nn
 \pa_\ald\rightarrow b_\ald(z)&\spa&\theta^\ald\rightarrow c^\ald(z)
\eea
in the differential operator realization $\{\tilde{J}_A\}$ and subsequently
to add anomalous terms
\bea
 F^{\mbox{{\footnotesize anom}}}_\al(\g(z),c(z),\pa\g(z),\pa c(z))&=&
  \pa\g^\beta(z)F_{\al\beta}(\g(z),c(z))+\pa c^\betad(z)F_{\al\betad}
  (\g(z),c(z))\nn
 f^{\mbox{{\footnotesize anom}}}_\ald(\g(z),c(z),\pa\g(z),\pa c(z))&=&
  \pa\g^\beta(z)f_{\ald\beta}(\g(z),c(z))+\pa c^\betad(z)f_{\ald\betad}
  (\g(z),c(z))
\eea
to the lowering operators $F_\al(z)$ and $f_\ald(z)$, respectively.
We are left with the following form of the free field realization
of the affine current superalgebra
\bea
 H_i(z)&=&:V_i^\beta(\g(z),c(z))\beta_\beta(z):+:V_i^\betad(\g(z),c(z))
  b_\betad(z):+\kvt\pa\var_i(z)\nn
 E_\al(z)&=&:\vab(\g(z),c(z))\beta_\beta(z):+:V_\al^\betad(\g(z),c(z))
  b_\betad(z):\nn
 F_\al(z)&=&:\vmab(\g(z),c(z))\beta_\beta(z):+:V_{-\al}^\betad(\g(z),c(z))
  b_\betad(z):+\kvt\pa\var_j(z)\paj(\g(z),c(z))\nn
 &+&\pa\g^\beta(z)F_{\al\beta}(\g(z),c(z))+\pa c^\betad(z)F_{\al\betad}
  (\g(z),c(z))\nn
 e_\ald(z)&=&:V_\ald^\beta(\g(z),c(z))\beta_\beta(z):
  +:V_\ald^\betad(\g(z),c(z))b_\betad(z):\nn
 f_\ald(z)&=&:V_{-\ald}^\beta(\g(z),c(z))\beta_\beta(z):+
  :V_{-\ald}^\betad(\g(z),c(z))b_\betad(z):
  +\kvt\pa\var_j(z)P_{-\ald}^j(\g(z),c(z))\nn
 &+&\pa\g^\beta(z)f_{\ald\beta}(\g(z),c(z))+\pa c^\betad(z)f_{\ald\betad}
  (\g(z),c(z))
\label{Wakimoto}
\eea
Now, the obvious task is to work out unique solutions for the anomalous terms.
This we will do in the following, and the result is one of the main new
results in this paper since it concludes the (generalized) Wakimoto free
field realization of affine current superalgebras (\ref{Wakimoto}).

A comparison of the free field realization (\ref{Wakimoto}) with (\ref{JAJB})
yields a set of quantum polynomial identities, listed in 
(\ref{quant}) in Appendix A.
Among those identities are
\bea 
 k\kappa_{\al,-\beta}&=&-\pa_\sigma V_\al^\gdd\pa_\gdd V_{-\beta}^\sigma
  +\pa_\sigmad V_\al^\gdd\pa_\gdd V_{-\beta}^\sigmad
  +V_\al^\gdd F_{\beta\gdd}\nn
 k\kappa_{\al,-\betad}&=&-\pa_\sigmadd V_\al^\gdd
  \pa_\gdd V_{-\betad}^\sigmadd+V_\al^\gdd f_{\betad\gdd}\nn
 k\kappa_{\ald,-\beta}&=&-\pa_\sigmadd V_\ald^\gdd\pa_\gdd V_{-\beta}^\sigmadd
  +V_\ald^\gdd F_{\beta\gdd}\nn
 k\kappa_{\ald,-\betad}&=&-\pa_\sigma V_\ald^\gdd\pa_\gdd V_{-\betad}^\sigma
  +\pa_\sigmad V_\ald^\gdd\pa_\gdd V_{-\betad}^\sigmad
  +V_\ald^\gdd f_{\betad\gdd}
\eea
following from the OPE's $E_\al F_\beta$, $E_\al f_\betad$, $e_\ald F_\beta$
and $e_\ald f_\betad$, respectively.
Hence, in order 
to determine the anomalous terms, we must show that $\vpp$ is invertible and
find its inverse. However, this follows immediately from (\ref{pol})
\ben
 \vpp(\g,c)={[B(C(\g,c))]_+}^+=B(\cpp(\g,c))
\een
and we may read off (\ref{Ber})
\bea
 (\vpp(\g,c))^{-1}&=&(B(\cpp(\g,c)))^{-1}\nn
 &=&\sum_{n\geq0}\frac{1}{(n+1)!}(\cpp(\g,c))^n
\eea
Thus we have
\bea
 F_{\al\betadd}(\g,c)&=&k\left[(\vpp(\g,c))^{-1}\right]_\betadd^\al
  \kappa_{\al,-\al}\nn
 &+&\left[(\vpp(\g,c))^{-1}\right]_\betadd^\mu\left(
  \pa_\sigma V_\mu^\gdd(\g,c)\pa_\gdd V_{-\al}^\sigma(\g,c)
  -\pa_\sigmad V_\mu^\gdd(\g,c)
  \pa_\gdd V_{-\al}^\sigmad(\g,c)\right)\nn
 &+&\left[(\vpp(\g,c))^{-1}\right]_\betadd^\mud\pa_\sigmadd
  V_\mud^\gdd(\g,c)\pa_\gdd V_{-\al}^\sigmadd(\g,c)\nn
 f_{\ald\betadd}(\g,c)&=&k\left[(\vpp(\g,c))^{-1}\right]_\betadd^\ald
  \kappa_{\ald,-\ald}\nn
 &+&\left[(\vpp(\g,c))^{-1}\right]_\betadd^\mu
 \pa_\sigmadd
  V_\mu^\gdd(\g,c)\pa_\gdd V_{-\ald}^\sigmadd(\g,c)\nn
 &+&\left[(\vpp(\g,c))^{-1}\right]_\betadd^\mud\left(\pa_\sigma
  V_\mud^\gdd(\g,c)\pa_\gdd V_{-\ald}^\sigma(\g,c)-\pa_\sigmad
  V_\mud^\gdd(\g,c)\pa_\gdd V_{-\ald}^\sigmad(\g,c)\right)
\eea
where we have used that $\kappa_{\aldd,-\betadd}=0$ unless $\aldd=\betadd$.
Note that there are no summations over $\al$ and $\ald$ in the terms
proportional to $k$. In particular for $\al$ or $\ald$ a simple root,
we find (using the common notation $F_{\aldd\betadd}$ for the anomalous parts
when $\aldd$ is a general positive root) 
\ben
 F_{\aldd_i\betadd}(\g,c)
  =\hf\delta_{\aldd_i,\betadd}\left((2k+\hn)\kappa_{\aldd_i,-\aldd_i}
  -A_{ii}\right)
\label{anomsimple}
\een
which is seen to be a constant, independent of $\g$ and $c$. Here we have used
that
\ben
 \pa_\sigma V_{\aldd_i}^\gdd(\g,c)\pa_\gdd V_{-\aldd_j}^\sigma(\g,c)
  -\pa_\sigmad V_{\aldd_i}^\gdd(\g,c)\pa_\gdd V_{-\aldd_j}^\sigmad(\g,c)
  =\hf\delta_{ij}\left(
  \hn\kappa_{\aldd_i,-\aldd_i}-A_{ii}\right)
\een

\section{Screening Currents}

A screening current has conformal weight 1 and has the property
that the singular part of the OPE with an affine current is a total 
derivative. These properties ensure that integrated
screening currents (screening charges) may be inserted into correlators
without altering the conformal or affine Ward identities. This in turn makes 
them very useful in construction of correlators, see e.g. 
\cite{DF,BF,Dot,PRY1,PRY2}.
The best known screening currents \cite{FF,BMP,Ito1,Ku,ATY,FFR,deBF,JR1,PRY3} 
are the ones of the first kind in standard bosonic affine current algebra.
Screening currents of the second second however \cite{BO1,Ito1,JR1,PRY3}, 
involve non-integer powers
of the ghost fields and therefore have been less studied. The techniques for
handling such objects have been developed in \cite{PRY1,PRY2}.

To the best of our knowledge, in the case of superalgebras the only known 
screening currents are the ones for $osp(1|2)$ and $sl(2|1)$. 
In the case of $osp(1|2)$, the screening current 
of the first kind is due to Bershadsky and Ooguri \cite{BO2} while the 
screening current of the second kind is due to Ennes {\em et al} \cite{ERdeS}.
In the case of $sl(2|1)$, only screening currents of the first kind are
known and are due to Ito \cite{Ito3}, and only for one choice of simple
roots, see Section 7.
In this section we will provide universal expressions for screening 
currents $s_{\aldd_j}(w)$ of the first kind for generic affine current 
superalgebras, thus presenting one of the main new results in this paper. 
Namely, we find the screening currents to be of the form
\bea
 s_{\aldd_j}(w)&=&:\left(S_{\aldd_j}^\sigma(\g(w),c(w))\beta_\sigma(w)+
  S_{\aldd_j}^\sigmad(\g(w),c(w))b_\sigmad(w)\right):
  :e^{-\aldd_j(H\cdot\var(w))/\kvt}:\nn
 &=&:\left(S_{\aldd_j}^\sigma(\g(w),c(w))\beta_\sigma(w)+
  S_{\aldd_j}^\sigmad(\g(w),c(w))b_\sigmad(w)\right)::e^{-\var_j(w)/(\kappa_{
  \aldd_j,-\aldd_j}\kvt)}:
\label{screen}
\eea 
where
\ben
 \pa\var_i(z)\La\cdot\var(w)=\frac{\La_i}{z-w}\spa \pa\var_i(z)\beta(
  H\cdot\var(w))=\frac{\beta(H_i)}{z-w}
\een 
and to produce the following total derivatives
\bea
 H_i(z)s_{\aldd_j}(w)&=&0\nn
 E_\beta(z)s_{\aldd_j}(w)&=&0\nn
 F_\beta(z)s_{\aldd_j}(w)&=&\frac{\pa}{\pa w}\left(
  \frac{(-1)^{p(\aldd_j)+1}\kappa_{\aldd_j,-\aldd_j}t}{z-w}
  Q_{-\beta}^{-\aldd_j}(\g(w),c(w)):e^{-\aldd_j(H\cdot\var(w))/\kvt}:\right)\nn
 e_\betad(z)s_{\aldd_j}(w)&=&0\nn
 f_\betad(z)s_{\aldd_j}(w)&=&\frac{\pa}{\pa w}\left(
  \frac{(-1)^{p(\aldd_j)+1}\kappa_{\aldd_j,-\aldd_j}t}{z-w}
  Q_{-\betad}^{-\aldd_j}(\g(w),c(w)):e^{-\aldd_j(H\cdot\var(w))/\kvt}:
  \right)\nn
 T(z)s_{\aldd_j}(w)&=&\frac{\pa}{\pa w}\left(\frac{1}{z-w}
  s_{\aldd_j}(w)\right)
\label{screender}
\eea
Utilizing the classical polynomial identities (\ref{Sclass}),
the proof is straightforward for the raising operators $E_\beta$ and 
$e_\betad$,
for the Cartan operator $H_i$ and for the energy-momentum tensor $T$. The last 
identity merely shows that indeed $s_{\aldd_j}(w)$ is a conformal primary
field with weight 1. For the lowering operators, comparisons of the two
sides in (\ref{screender}) yield the following consistency conditions
\bea
 t\kappa_{\aldd_j,-\aldd_j}Q_{-\betadd}^{-\aldd_j}&=&-(-1)^{p(\aldd_j)
  (1-p(\betadd))}\left(S_{\aldd_j}^\sigmadd F_{\betadd\sigmadd}
  +A_{ij}S_{\aldd_j}^\sigmadd \pa_\sigmadd P_{-\betadd}^i\right)\nn
 &+&(-1)^{p(\aldd_j)}\pa_\sigma V_{-\betadd}^\gdd\pa_\gdd S_{\aldd_j}^\sigma
  -(-1)^{p(\betadd)}\pa_\sigmad V_{-\betadd}^\gdd\pa_\gdd 
  S_{\aldd_j}^\sigmad\nn
 t\kappa_{\aldd_j,-\aldd_j}\pa_\gdd Q_{-\betadd}^{-\aldd_j}&=&
  (-1)^{p(\aldd_j)(1-p(\betadd)-p(\gdd))}\left(-S_{\aldd_j}^\sigma\pa_\gdd
  F_{\betadd\sigma}-(-1)^{p(\gdd)}S_{\aldd_j}^\sigmad\pa_\gdd
  F_{\betadd\sigmad}\right.\nn
 &-&\left.A_{ij}S_{\aldd_j}^\sigmadd\pa_\sigmadd\pa_\gdd
  P_{-\betadd}^i+S_{\aldd_j}^\sigmadd\pa_\sigmadd F_{\betadd\gdd}\right)\nn
 &+&(-1)^{p(\aldd_j)}\pa_\sigma\pa_\gdd V_{-\betadd}^\mudd\pa_\mudd 
  S_{\aldd_j}^\sigma-(-1)^{p(\betadd)+p(\gdd)}\pa_\sigmad\pa_\gdd
  V_{-\betadd}^\mudd\pa_\mudd S_{\aldd_j}^\sigmad
\label{Scond}
\eea 
They are easily verified for $\betadd$ a simple root
$\betadd=\aldd_i$, using that
\ben
 \pa_\sigma V_{-\aldd_i}^\gdd\pa_\gdd S_{\aldd_j}^\sigma-\pa_\sigmad
  V_{-\aldd_i}^\gdd\pa_\gdd S_{\aldd_j}^\sigmad=\hf\delta_{ij}
  (-1)^{p(\aldd_i)}\left(\hn\kappa_{\aldd_i,-\aldd_i}-A_{ii}\right)
\een
In the case of a non-simple root $\betadd$, we have proven the conditions 
(\ref{Scond}) by induction in addition of roots using various classical 
and quantum polynomial identities. The strategy is fairly
straightforward, though very tedious. First one eliminates
terms involving $F_{\betadd\sigmadd}$, using in particular
the recursion relations (\ref{rec1}) and (\ref{rec2}) 
expressing e.g. ${f_{\aldd,\aldd'}}^\betadd 
F_{\betadd\sigmadd}$ in terms of polynomials with indices $\aldd$ and
$\aldd'$. By induction assumption, such polynomials do satisfy (\ref{Scond}),
so by substitution we may get rid of terms involving derivatives of
the anomalous polynomials. Next one eliminates terms involving 
$F_{\aldd\sigmadd}$, and we are
left with a set of relations in $V$, $P$, and $S$. (The $Q$'s in 
(\ref{Scond}) may be expressed in terms of $P$'s and $S$'s  
(\ref{Sclass2})). The proof is then concluded by virtue of the classical 
polynomial identities. Throughout we may act on the various identities with 
appropriate differential operators in order to derive further identities.

Note that the numbers of fermionic and bosonic screening currents depend
on the underlying Lie superalgebra and the choice of a set of simple roots
in that.  

\section{Primary Fields}

The final main new result reported in this paper is the explicit 
construction in this section of primary fields for arbitrary representations, 
integral or non-integral. 
We find it particularly convenient to replace the traditional 
multiplet of primary fields (which generically would be infinite for 
non-integrable representations) by a generating function for that, namely 
the primary field $\phi_\La(w,x,\theta)$ which must satisfy
\bea
 J_A(z)\phi_\La(w,x,\theta)&=&\frac{-J_A(x,\theta,\pa,\La)}{z-w}
  \phi_\La(w,x,\theta)\nn
 T(z)\phi_\La(w,x,\theta)&=&\frac{\Delta(\phi_\La)}{(z-w)^2}
  \phi_\La(w,x,\theta)
  +\frac{1}{z-w}\pa\phi_\La(w,x,\theta)
\label{primdef}
\eea
Here $J_A(z)$ are the affine currents, whereas 
$J_A(x,\theta,\pa,\La)$ are the differential operator realizations given in 
(\ref{ketdef}), (\ref{tilde}), (\ref{superadjoint}), (\ref{defVP}) 
and (\ref{pol}).
We shall find the result in the form
\bea
 \phi_\La(w,x,\theta)&=&\phi_\La'(\g(w),c(w),x,\theta)V_\La(w) \nn
 V_\La(w)&=&:e^{\frac{1}{\kvt}\La\cdot\var(w)}:\nn
\phi_\La'(\g(w),c(w),0,0)&=&1
\label{primans}
\eea
Indeed, such a field is conformally primary and has conformal dimension
$\Delta(\phi_\La)=\frac{1}{2t}(\La,\La+2\rho)$. In order to comply with
(\ref{primdef}) for $J_A=H_i$, $\phi_\La'$ is seen to 
be supersymmetric in $(x,\theta)$ and $(\g(w),c(w))$. 
Below we shall demonstrate this by explicit 
construction. Due to the fact that the anomalous parts of 
$F_\al(z)$ and $f_\ald(z)$ do not give singular contributions
when contracting with $\phi_\La'$, we are left with the following sufficient
conditions on $\phi_\La'(\g(w),c(w),x,\theta)$, a pair for each $\aldd>0$
\bea
 (-1)^{p(\aldd)}\left(V_\aldd^\beta(\g,c)\pa_{\g^\beta}
  +V_\aldd^\betad(\g,c)\pa_{c^\betad}\right)\phi_\La'&=&
 \left(V_{-\aldd}^\beta(x,\theta)\pa_{x^\beta}+
  V_{-\aldd}^\betad(x,\theta)\pa_{\theta^\betad}\right)\phi_\La'\nn
 &+&\La_jP_{-\aldd}^j(x,\theta)\phi_\La'\nn
 \left(V_\aldd^\beta(x,\theta)\pa_{x^\beta}+V_\aldd^\betad(x,\theta)
  \pa_{\theta^\betad}\right)\phi_\La'&=&
 \left(V_{-\aldd}^\beta(\g,c)\pa_{\g^\beta}+
  V_{-\aldd}^\betad(\g,c)\pa_{c^\betad}\right)\phi_\La'\nn
 &+&\La_jP_{-\aldd}^j(\g,c)\phi_\La'
\label{primsuff}
\eea
where the sign factor is due to (\ref{tilde}) and (\ref{superadjoint}).
Further, one can use the classical polynomial 
identities (\ref{rec1}) to reduce the numbers of necessary conditions. 
Let us assume that $\aldd>0$ is
non-simple such that there exist $\betadd>0$ and $\gdd>0$ satisfying
${f_{\betadd,\gdd}}^\aldd\neq0\neq{f_{-\betadd,-\gdd}}^{-\aldd}$. We may
then multiply the left hand sides by ${f_{\betadd,\gdd}}^\aldd$
and the right hand sides by $\frac{{f_{\betadd,\gdd}}^\aldd}{
{f_{-\betadd,-\gdd}}^{-\aldd}}{f_{-\betadd,-\gdd}}^{-\aldd}$. By virtue
of (\ref{rec1}) it then follows that ${f_{\betadd,\gdd}}^{\aldd}/
{f_{-\betadd,-\gdd}}^{-\aldd}=-1$, in accordance with (\ref{fsymm2}). In 
conclusion, there are only $2r$ sufficient conditions a primary field must 
satisfy 
\bea
 (-1)^{p(\aldd_i)}
  \left(V_{\aldd_i}^\beta(\g,c)\pa_{\g^\beta}+V_{\aldd_i}^\betad(\g,c)
  \pa_{c^\betad}\right)\phi_\La'&=&
  \left(V_{-\aldd_i}^\beta(x,\theta)\pa_{x^\beta}+
  V_{-\aldd_i}^\betad(x,\theta)\pa_{\theta^\betad}\right)\phi_\La'\nn
 &+&\La_jP_{-\aldd_i}^j(x,\theta)\phi_\La'\nn
 \left(V_{\aldd_i}^\beta(x,\theta)\pa_{x^\beta}+V_{\aldd_i}^\betad(x,\theta)
  \pa_{\theta^\betad}\right)\phi_\La'&=&
 \left(V_{-\aldd_i}^\beta(\g,c)\pa_{\g^\beta}+
  V_{-\aldd_i}^\betad(\g,c)\pa_{c^\betad}\right)\phi_\La'\nn
 &+&\La_jP_{-\aldd_i}^j(\g,c)\phi_\La'
\label{primeq} 
\eea
where
\ben
 P_{-\al_i}^j(x,\theta)=\delta_i^jx^{\al_j}\spa P_{-\ald_i}^j(x,\theta)
  =\delta_i^j\theta^{\ald_j}
\een
It seems very hard to solve this set of partial differential equations 
directly. However, in \cite{PRY3} an alternative way to obtain the primary 
fields was developed in the case of purely bosonic
affine current algebras. The analogous construction for affine current
superalgebras goes as follows.

First we directly construct primary fields for each basis or
fundamental representation
$M_{\La^k}$. Such representation spaces are finite dimensional modules 
and $\phi'_{\La^k}(\g(w),c(w),x,\theta)$ will be polynomial in $\g(w)$,
$c(w)$, $x$ and $\theta$.
Then finally, for a general representation with highest weight 
$\La=\la_k\La^k$ (see (\ref{Dynkin}))
we use (\ref{primeq}) to immediately verify that
\ben
 \phi'_{\La}(\g(w),c(w),x,\theta)=\prod_{k=1}^r{[}\phi'_{\La^k}
  (\g(w),c(w),x,\theta){]}^{\la_k}
\label{ansatz}
\een
We emphasize here that the labels $\la_k$ may be 
non-integers, as is required for degenerate representations.
We proceed to explain how to construct the building blocks
\ben
 \phi'_{\La^k}(\g(w),c(w),x,\theta)
\een

The strategy goes as follows. First we concentrate on the case $w=0$ where
the object reduces to
\ben
 \phi'_{\La^k}(\g_0,c_0,x,\theta)
\label{phimgo}
\een
when acting on the highest weight state $\ket{\La^k}$.
$\g_0$ and $c_0$ are the zero modes in the mode expansions
\bea
 \g(w)=\sum_n \g_n w^{-n}&\spa&\beta(w)=\sum_n\beta_nw^{-n-1}\nn
 c(w)=\sum_n c_n w^{-n}&\spa&b(w)=\sum_nb_nw^{-n-1}
\eea
Conformal covariance requires $\phi'_{\La^k}(\g(w),c(w),x,\theta)$ 
to be obtained just
by replacing $\g_0$ by $\g(w)$ and $c_0$ by $c(w)$. 
The function (\ref{phimgo}) in turn is obtained from
\ben
 \ket{\La^k,x,\theta}=G_-(x,\theta)\ket{\La^k}=\phi'_{\La^k}(\g_0,c_0,x,\theta)
  \ket{\La^k}
\een
Indeed, it is a consequence of the formalism, that the primary
field constructed this way will satisfy the OPE (\ref{primdef}). The
construction is now simply achieved by expanding the state
$\ket{\La^k,x,\theta}$ on an appropriate basis which is convenient to obtain
using the free field realization.

Let the orthonormal basis elements in the $k$'th fundamental highest 
weight module $M_{\La^k}$ be 
denoted $\{\ket{U,\La^k}\}$ such that the identity operator may be written
\ben
 I=\sum_U\ket{U,\La^k}\bra{U,\La^k}
\label{identity}
\een
The state $\ket{\La^k,x,\theta}$ may then be written
\ben
 \ket{\La^k,x,\theta}=\sum_U\ket{U,\La^k}\bra{U,\La^k}\La^k,x,\theta\rangle
\een
One of the basis vectors will always be taken to be the highest weight 
vector $\ket{\La^k}$ itself. 

A particular basis vector will be of the form
\ben
 \ket{U,\La^k}\sim 
  F_{\betadd_1^{(U)}}...F_{\betadd_{n(U)}^{(U)}}\ket{\La^k}\spa
  n(U)=n^0(U)+n^1(U)
\label{basisdef}
\een
and the expansion of $\ket{\La^k,x,\theta}$ will be
\bea
 \ket{\La^k,x,\theta}&=&\sum_U\frac{1}{
  \bra{\La^k}E_{\betadd_{n(U)}^{(U)}}...E_{\betadd_1^{(U)}}
  F_{\betadd_1^{(U)}}...F_{\betadd_{n(U)}^{(U)}}\ket{\La^k}}\nn
 &\cdot&F_{\betadd_1^{(U)}}...F_{\betadd_{n(U)}^{(U)}}\ket{\La^k}
 \bra{\La^k}E_{\betadd_{n(U)}^{(U)}}...E_{\betadd_1^{(U)}}\ket{\La^k,x,\theta}
\label{ILax}
\eea
Here we have used the common notation $F_\betadd$ ($E_\betadd$) for the 
lowering (raising) generators corresponding to the positive root $\betadd$.
The parameters $n^0(U)$ and $n^1(U)$ denote the numbers of even and odd
generators (respectively) appearing in the expression (\ref{basisdef}).
For each term in the sum (\ref{ILax})
we treat the two factors differently. First consider 
the second factor. We may use the differential operator realizations to write
\bea
 &&\bra{\La^k}E_{\betadd_{n(U)}^{(U)}}...
  E_{\betadd_1^{(U)}}\ket{\La^k,x,\theta}\nn
 &=&(-1)^{n^1(U)(n^1(U)-1)/2+n(U)}E_{\betadd_1^{(U)}}(x,\theta,\pa,\La^k)
  ...E_{\betadd_{n(U)}^{(U)}}
  (x,\theta,\pa,\La^k)\bra{\La^k}\La^k,x,\theta\rangle\nn
 &=&(-1)^{n^1(U)(n^1(U)+1)/2}U(x,\theta,\pa/\pa x,\pa/\pa \theta,\La^k)
\label{Ufunc}
\eea
where
\bea
 &&U(x,\theta,\pa/\pa x,\pa/\pa \theta,\La^k)\nn
 &=&\tilde{F}_{\betadd_1^{(U)}}
  (x,\theta,\pa,\La^k)...\tilde{F}_{\betadd_{n(U)}^{(U)}}
  (x,\theta,\pa,\La^k)\cdot1\nn
 &=&\left[V^{\gdd_1}_{-\betadd_1^{(U)}}(x,\theta)\pa_{\gdd_1}+
  \La^k(H_j)P^j_{-\betadd_1^{(U)}}(x,\theta)\right]...\nn
 &\cdot&\left[V^{\gdd_{n(U)-1}}_{-\betadd_{n(U)-1}^{(U)}}(x,\theta)
  \pa_{\gdd_{n(U)-1}}+\La^k(H_j)P^j_{-\betadd_{n(U)-1}^{(U)}}(x,\theta)\right]
  \La^k(H_j)P^j_{-\betadd_{n(U)}^{(U)}}(x,\theta)
\label{primpol}
\eea
In the last step in (\ref{Ufunc}) we used that clearly
\ben
 \bra{\La^k}\La^k,x,\theta\rangle \equiv 1
\een
Actually, the function $U$ is independent of $\pa/\pa x$ and $\pa/\pa \theta$
since the differentiations may easily be carried out.
 
In the first factor in (\ref{ILax}) 
\ben
  F_{\betadd_1^{(U)}}...F_{\betadd_{n(U)}^{(U)}}\ket{\La^k}
\een
we use the free field realizations. The state $\ket{\La^k}$ is a vacuum for
the $\beta,\g$ and the $b,c$ systems, 
so it is annihilated by $\g_n,\ n\geq 1$ and $\beta_n,\ n\geq 0$, and by
$c_n,\ n\geq1$ and $b_n,\ n\geq0$. 
The $F_\betadd$'s are the zero modes of the affine currents (\ref{identif}).
It follows that only $\g_0$'s, $\beta_0$'s, $c_0$'s and $b_0$'s need be 
considered. Also the anomalous terms will not contribute, and we obtain
\bea
 &&F_{\betadd_1^{(U)}}...F_{\betadd_{n(U)}^{(U)}}\ket{\La^k}\nn
 &=&\left[V^{\g_1}_{-\betadd_1^{(U)}}(\g_0,c_0)\beta_{\g_1,0}+
  V^{\gd_1}_{-\betadd_1^{(U)}}(\g_0,c_0)b_{\gd_1,0}+
  \La^k(H_j)P^j_{-\betadd_1^{(U)}}(\g_0,c_0)\right]...\nn
 &\cdot&\left[V^{\g_{n(U)-1}}_{-\betadd_{n(U)-1}^{(U)}}(\g_0,c_0)
  \beta_{\g_{n(U)-1},0}+V^{\gd_{n(U)-1}}_{-\betadd_{n(U)-1}^{(U)}}(\g_0,c_0)
  b_{\gd_{n(U)-1},0}+\La^k(H_j)
  P^j_{-\betadd_{n(U)-1}^{(U)}}(\g_0,c_0)\right]\nn
 &\cdot&\La^k(H_j)P^j_{-\betadd_{n(U)}^{(U)}}(\g_0,c_0)\ket{\La^k}\nn
 &=&U(\g_0,c_0,\beta_0,b_0,\La^k)\ket{\La^k}
\eea
As before, the function $U$ is independent of $\beta_0$ and $b_0$.
By the remarks above this completes the construction in general:
\bea 
 &&\phi_{\La^k}'(\g(w),c(w),x,\theta)\nn
 &=&\sum_U\frac{(-1)^{n^1(U)(n^1(U)+1)/2}}{
  \bra{\La^k}E_{\betadd_{n(U)}^{(U)}}...E_{\betadd_1^{(U)}}
  F_{\betadd_1^{(U)}}...F_{\betadd_{n(U)}^{(U)}}\ket{\La^k}} 
  U(\g(w),c(w),\La^k)U(x,\theta,\La^k)\nn
 &=&\sum_U\frac{(-1)^{n^1(U)(n^1(U)-1)/2}}{
  \bra{\La^k}E_{\betadd_{n(U)}^{(U)}}...E_{\betadd_1^{(U)}}
  F_{\betadd_1^{(U)}}...F_{\betadd_{n(U)}^{(U)}}\ket{\La^k}}
 U(x,\theta,\La^k)U(\g(w),c(w),\La^k)
\label{primgen}
\eea
Explicit expressions for the $V$'s and the $P$'s have already been provided 
(\ref{pol}).

It remains to account in detail for how to obtain a convenient basis for the 
fundamental representations. This part will depend on the Lie superalgebra
in question, and we anticipate that there are no problems in writing down
such a basis. In the case of standard affine $SL(N)$ current
algebra, this has been done in \cite{PRY3}. See Section 7 for explicit
constructions of primary fields in particular cases of affine current
superalgebras.

\section{Examples}

In this section we shall discuss several examples of the general constructions
provided in previous sections, in order to illustrate how powerful these
constructions are and to compare with results known in the literature.
  
\subsection{Chevalley Generators}

In \cite{Ito2} Ito has considered realizations of Chevalley generators of both 
classical Lie superalgebras and of affine current superalgebras. Actually,
he considers also the trivial extensions to any raising generator
(in the $\{ \tilde{J}_A\}$ basis, or lowering generators in the $\{ J_A\}$
basis). It is not difficult to verify that our similar results reduce to 
those of Ito. Indeed, we find the following key objects in the 
differential operator realizations (see (\ref{tilde}), (\ref{defVP}) and
(\ref{pol})) 
\bea
 V_\aldd^\betadd(x,\theta)&=&\delta_\aldd^\betadd+\sum_{n\geq1}\frac{B_n}{n!}
  (-1)^n{f_{\betadd_1,\aldd}}^{\gdd_1}{f_{\betadd_2,\gdd_1}}^{\gdd_2}...
  {f_{\betadd_n,\gdd_{n-1}}}^{\betadd}\nn
 &\cdot&\left(x^{\beta_1}\delta_{\beta_1}^{\betadd_1}+\theta^{\betad_1}
  \delta_{\betad_1}^{\betadd_1}\right)...
  \left(x^{\beta_n}\delta_{\beta_n}^{\betadd_n}+\theta^{\betad_n}
  \delta_{\betad_n}^{\betadd_n}\right)\nn
 V_i^\betadd(x,\theta)&=&-\betadd(H_i)\left(x^\beta\delta_\beta^\betadd 
  +\theta^\betad\delta_\betad^\betadd\right)\nn
 V_{-\aldd_i}^\betadd(x,\theta)&=&{f_{\mudd,-\aldd_i}}^\betadd\left(x^\mu
  \delta_\mu^\mudd+\theta^\mud\delta_\mud^\mudd\right)
  +\sum_{n\geq1}\frac{B_n}{n!}
  \betadd_1(H_i){f_{\betadd_2,\betadd_1}}^{\gdd_1}
  {f_{\betadd_3,\gdd_1}}^{\gdd_2}...{f_{\betadd_n,\gdd_{n-2}}}^{\betadd}\nn
 &\cdot&(x^{\al_i}\delta_{\al_i,\aldd_i}+\theta^{\ald_i}\delta_{\ald_i,
  \aldd_i})\left(x^{\beta_1}\delta_{\beta_1}^{\betadd_1}+\theta^{\betad_1}
  \delta_{\betad_1}^{\betadd_1}\right)...
  \left(x^{\beta_n}\delta_{\beta_n}^{\betadd_n}+\theta^{\betad_n}
  \delta_{\betad_n}^{\betadd_n}\right)\nn
 P_{-\aldd_i}^j(x,\theta)&=&\delta_i^j\left(x^{\al_j}\delta_{\al_j}^{\aldd_j}
  +\theta^{\ald_j}\delta_{\ald_j}^{\aldd_j}\right)
\eea 
This is in accordance with \cite{Ito2}.
The expression for $V_{-\aldd_i}^\betadd(x,\theta)$ is a simple reduction of 
the following non-trivial rewriting of $V_{-\aldd}^\betadd(x,\theta)$
(\ref{pol}), valid for all positive roots $\aldd$
\bea
 V_{-\aldd}^\betadd(x,\theta)&=&\sum_{n\geq1}\sum_{\betadd_i>0}\sum_{l=0}^{
  n-m(-\aldd,\betadd_1,...,\betadd_n)}\frac{B_l}{l!(n-l)!}
  {f_{\betadd_1,-\aldd}}^{\gdd_1}{f_{\betadd_2,\gdd_1}}^{\gdd_2}...
  {f_{\betadd_n,\gdd_{n-1}}}^{\betadd}\nn
 &\cdot&\left(x^{\beta_1}\delta_{\beta_1}^{\betadd_1}+\theta^{\betad_1}
  \delta_{\betad_1}^{\betadd_1}\right)...
  \left(x^{\beta_n}\delta_{\beta_n}^{\betadd_n}+\theta^{\betad_n}
  \delta_{\betad_n}^{\betadd_n}\right)
\label{rewriteV}
\eea
where $m(-\aldd,\betadd_1,...,\betadd_n)$ is defined for a given sequence
of roots $(-\aldd,\betadd_1,...,\betadd_m,...,\betadd_n)$ 
as the minimum integer
for which $-\aldd+\betadd_1+...+\betadd_m>0$.
Our proof of the rewriting is based on the lemma stating that in any (formal) 
expansion of the form
\ben
 \sum_{s\geq0}\left(-\sum_{n\geq1}b_nx^n\right)^s=\sum_{n\geq0}a_nx^n
\een
the following recursion relation is valid
\bea
 a_0&=&1\nn
 a_n&=&-\sum_{l=0}^{n-1}b_{n-l}a_l\spa \mbox{for}\ n>0
\eea
Proofs of (the bosonic equivalence of) (\ref{rewriteV}) and of the lemma
may be found in \cite{JR1}. 
In the affine current superalgebras, Ito \cite{Ito2} 
works out the anomalous terms only for Chevalley generators
(corresponding to simple roots). His result is in complete agreement
with ours given in (\ref{anomsimple}).

\subsection{Case of $OSp(1|2)$}

The classical Lie superalgebra $osp(1|2)$, isomorphic to $B(0,1)$,
consists of 3 bosonic generators
$E$, $H$ and $F$, and 2 fermionic generators $e$ and $f$. This corresponds to 
the fact that there are exactly 2 positive roots, namely one odd simple
root $\ald$ and one even non-simple root $\al=2\ald$. Thus, the Weyl vector
becomes $\rho=\hf\ald$. Let us impose the normalization 
condition $\al^2=2$, or 
equivalently $\hn=3/2$ (following from Freudenthal-de Vries (super-)strange
formula), and also use the convention of Kac setting $A_{11}=2$.
This leads to some slightly unconventional normalizations of the 
Cartan-Killing form
\ben
 \kappa_{\ald,-\ald}=\kappa_{\al,-\al}=4\spa G_{11}=8
\een
and of the structure coefficients
\bea
 {f_{1,\pm\ald}}^{\pm\ald}=\pm2&\spa&{f_{1,\pm\al}}^{\pm\al}=\pm4\nn
 {f_{\ald,-\ald}}^1=1&\spa&{f_{\al,-\al}}^1=2\nn
 {f_{\pm\ald,\pm\ald}}^{\pm\al}=\pm2&\spa&{f_{\pm\ald,\mp\al}}^{\mp\ald}=2
\eea
These parameters may be obtained from the more standard ones (see e.g. 
\cite{FY})
\bea
 \left[ J_3,j_\pm\right]=\pm\hf j_\pm&\spa&\left[ J_3,J_\pm\right]=\pm J_\pm\nn
 \left\{ j_+,j_-\right\}=2J_3&\spa&\left[ J_+,J_-\right]=2J_3\nn
 \left\{ j_\pm,j_\pm\right\}=\pm2J_\pm&\spa&\left[J_\pm,j_\mp\right]=-j_\pm
\eea
by the substitutions
\bea
 J_+=\hf E\spa&J_3=\frac{1}{4}H&\spa J_-=\hf F\nn
 j_+=\frac{1}{\sqrt{2}}e&\spa&j_-=\frac{1}{\sqrt{2}}f
\eea
The general results in previous sections are easily reduced in the case
of $OSp(1|2)$. We find the differential operator realizations 
\ben
 \left\{\begin{array}{l}
  \tilde{H}(x,\theta,\pa,\La)=-2\theta\pa_\theta-4x\pa_x+\La_1\\
   \tilde{e}(x,\theta,\pa)=\pa_\theta+\theta\pa_x\\
  \tilde{f}(x,\theta,\pa,\La)=-2x\pa_\theta-2\theta x\pa_x+\theta\La_1\\
  \tilde{E}(x,\theta,\pa)=\pa_x\\
   \tilde{F}(x,\theta,\pa,\La)=-4\theta x\pa_\theta-4x^2\pa_x+2x\La_1
 \end{array}\right.   \hspace{.5cm} \left\{
  \begin{array}{l}
 H(x,\theta,\pa,\La)=2\theta\pa_\theta+4x\pa_x-\La_1\\
   e(x,\theta,\pa,\La)=-2x\pa_\theta-2\theta x\pa_x+\theta\La_1\\
  f(x,\theta,\pa)=-\pa_\theta-\theta\pa_x\\
 E(x,\theta,\pa,\La)=4\theta x\pa_\theta+4x^2\pa_x-2x\La_1\\
   F(x,\theta,\pa)=-\pa_\al
 \end{array}\right.
\een
The Wakimoto free field realization is based on one pair of bosonic ghost
fields $(\beta,\g)$, one pair of fermionic ghost fields $(b,c)$ and
one bosonic scalar field $\var=\var_1$ satisfying $\var(z)\var(w)=G_{11}
\ln(z-w)=8\ln(z-w)$, and is found to be
\bea
 H(z)&=&-2:c(z)b(z):-4:\g(z)\beta(z):+\kvt\pa\var(z)\nn
 e(z)&=&b(z)+c(z)\beta(z)\nn
 f(z)&=&-2\g(z)b(z)-2c(z):\g(z)\beta(z):+\kvt c(z)\pa\var(z)+2(2k+1)\pa c(z)\nn
 E(z)&=&\beta(z)\nn
 F(z)&=&-4\g(z):c(z)b(z):-4:\g^2(z)\beta(z):+2\kvt\g(z)\pa\var(z)\nn
 &-&4(k+1)\pa c(z)c(z)+4k\pa\g(z)
\eea
This is in accordance with the literature \cite{BO2}. It is easily verified
that indeed (\ref{JAJB}) is satisfied. The Sugawara energy-momentum tensor
becomes
\ben
 T(z)=:\pa\g(z)\beta(z):+:\pa c(z)b(z):+\hf:\pa\var(z)\cdot\pa\var(z):
  -\frac{1}{2\kvt}\ald\cdot\pa^2\var(z)
\een
The screening current of the first kind is found to be
\ben
 s_\ald(w)=\left(c(w)\beta(w)-b(w)\right):e^{-\var(w)/(4\kvt)}:
\een
and is also known in the literature \cite{BO2}. It is not difficult to
show that the screening current of the second kind is \cite{ERdeS}
\ben
 \tilde{s}_\ald(w)=\left(c(w)\beta(w)-b(w)\right)\beta^{-(k+2)}(w)
  :e^{\kvt\var(w)/2}:
\een
and that it produces the following total derivatives
\bea
 F(z)\tilde{s}_\ald(w)&=&\frac{\pa}{\pa w}\left(\frac{4}{z-w}
  \left((k+1)c(w)\beta(w)-(k+2)b(w)\right)\beta^{-(k+3)}(w):e^{\kvt\var(w)/2}:
  \right)\nn
 f(z)\tilde{s}_\ald(w)&=&\frac{\pa}{\pa w}\left(\frac{2}{z-w}
  (-\beta^{-(k+2)}(w)):e^{\kvt\var(w)/2}:\right)
\eea
OPE's with the remaining affine current generators simply vanish. 
The primary field is found to be 
\ben
 \phi_\La(w,x,\theta)=\left(1+2\theta c(w)+4x\g(w)\right)^{\La_1/2}
  :e^{\La\cdot\var(w)/\kvt}:
\een
This result has also been obtained in \cite{ERdeS}, though based on different
normalization conventions for the Lie superalgebra parameters.

\subsection{Case of $OSp(2|2)\simeq SL(2|1)\simeq SL(1|2)$}

The Lie superalgebra $A(1,0)$, which is isomorphic to $osp(2|2)$, $C(2)$ and
$sl(1|2)\simeq sl(2|1)$, has rank $r=2$ and 3 positive roots, while
the dual Coxeter number is $\hn=1$ and dim$(\mbox{{\bf g}}_0)=$ 
dim$(\mbox{{\bf g}}_1)=4$.
First we choose the set of simple roots to consist of one even
simple root $\al_1$ and one odd simple root $\ald_2$. The remaining and
non-simple root $\ald=\al_1+\ald_2$ is then odd. 
The Weyl vector is $\rho=-\ald_2$. From the oscillator realization 
(see \cite{Tan})
\bea
 H_1&=&a_1^\dagger a_1-a_2^\dagger a_2\spa H_2=a_2^\dagger a_2+b^\dagger b\nn
 E_{\al_1}&=&a_1^\dagger a_2\hspace{1.3cm}\spa F_{\al_1}=a_2^\dagger a_1\nn
 e_{\ald_2}&=&a_2^\dagger b\hspace{1.5cm}\spa f_{\ald_2}=b^\dagger a_2\nn
 e_\ald&=&a_1^\dagger b\hspace{1.5cm}\spa f_\ald=b^\dagger a_1
\label{osc}
\eea
where $a_i^{(\dagger)}$ and $b^{(\dagger)}$ 
are fermionic and bosonic oscillators satisfying
\ben
 [b,b^\dagger]=1\spa\{a_i,a_j^\dagger\}=\delta_{ij}
  \spa [b^{(\dagger)},a_i^{(\dagger)}]=0
\een
we find the non-vanishing elements of the Cartan-Killing form to be 
\bea
 G_{11}=2\spa G_{12}&=&G_{21}=-1\spa G_{22}=0\nn
 \kappa_{\al_1,-\al_1}&=&\kappa_{\ald_2,-\ald_2}=\kappa_{\ald,-\ald}=1
\label{A10G}
\eea
such that the Cartan matrix is given by $A_{ij}=G_{ij}$. 
The remaining non-vanishing structure coefficients are found to be
\bea
 {f_{\al_1,-\al_1}}^1=1&\spa&{f_{\ald_2,-\ald_2}}^2=1\nn
 \ald(H_1)={f_{1,\ald}}^\ald=1&\spa&\ald(H_2)={f_{2,\ald}}^\ald=-1\nn
 {f_{\ald,-\ald}}^1={f_{\ald,-\ald}}^2=1
  &\spa&{f_{\pm\ald_2,\mp\ald}}^{\mp\al_1}=1\nn
 {f_{\pm\al_1,\pm\ald_2}}^{\pm\ald}=\pm1&\spa&
  {f_{\pm\al_1,\mp\ald}}^{\mp\ald_2}=\mp1
\label{A10f}
\eea

Now, the differential operator realization $\{\tilde{J}_A\}$ is worked out
to be
\bea
 \tilde{H}_1(x,\theta,\pa,\La)&=&-2x^{\al_1}\pa_{\al_1}+\theta^{\ald_2}
  \pa_{\ald_2}-\theta^\ald\pa_\ald+\La_1\nn
 \tilde{H}_2(x,\theta,\pa,\La)&=&x^{\al_1}\pa_{\al_1}+\theta^\ald\pa_\ald
  +\La_2\nn
 \tilde{E}_{\al_1}(x,\theta,\pa)&=&\pa_{\al_1}-\hf\theta^{\ald_2}\pa_\ald\nn
 \tilde{F}_{\al_1}(x,\theta,\pa,\La)&=&-x^{\al_1}x^{\al_1}\pa_{\al_1}
  +\left(\hf x^{\al_1}\theta^{\ald_2}-\theta^\ald\right)\pa_{\ald_2}\nn
 &-&\hf x^{\al_1}\left(\hf x^{\al_1}\theta^{\ald_2}+\theta^\ald\right)\pa_\ald
  +x^{\al_1}\La_1\nn
 \tilde{e}_{\ald_2}(x,\theta,\pa)&=&\pa_{\ald_2}+\hf x^{\al_1}\pa_\ald\nn
 \tilde{f}_{\ald_2}(x,\theta,\pa,\La)&=&\left(\hf x^{\al_1}\theta^{\ald_2}
  +\theta^\ald\right)\pa_{\al_1}+\hf\theta^{\ald_2}\theta^\ald\pa_\ald
  +\theta^{\ald_2}\La_2\nn
 \tilde{e}_{\ald}(x,\theta,\pa)&=&\pa_\ald\nn
 \tilde{f}_{\ald}(x,\theta,\pa,\La)&=&-x^{\al_1}\left(\hf x^{\al_1}
  \theta^{\ald_2}+\theta^\ald\right)\pa_{\al_1}-\theta^{\ald_2}\theta^\ald
  \pa_{\ald_2}\nn
 &+&\left(\hf x^{\al_1}\theta^{\ald_2}+\theta^\ald\right)\La_1-\left(
  \hf x^{\al_1}\theta^{\ald_2}-\theta^\ald\right)\La_2
\eea
The alternative realization $\{J_A\}$ is easily derived from this.

The (generalized) Wakimoto free field realization of the associated 
affine current superalgebra becomes
\bea
 H_1(z)&=&-2:\g(z)\beta(z):+:c(z)b(z):-:C(z)B(z):+\kvt\pa\var_1(z)\nn
 H_2(z)&=&:\g(z)\beta(z):+:C(z)B(z):+\kvt\pa\var_2(z)\nn
 E_{\al_1}(z)&=&\beta(z)-\hf c(z)B(z)\nn
 F_{\al_1}(z)&=&-:\g^2(z)\beta(z):+:\left(\hf\g(z)c(z)-C(z)\right)b(z):\nn
 &-&\hf:\g(z)\left(\hf\g(z)c(z)+C(z)\right)B(z):
  +\kvt\g(z)\pa\var_1(z)+(k-1/2)\pa\g(z)\nn
 e_{\ald_2}(z)&=&b(z)+\hf\g(z)B(z)\nn
 f_{\ald_2}(z)&=&:\left(\hf\g(z)c(z)+C(z)\right)\beta(z):+\hf c(z):C(z)B(z):\nn
 &+&\kvt c(z)\pa\var_2(z)+(k+1/2)\pa c(z)\nn
 e_\ald(z)&=&B(z)\nn
 f_\ald(z)&=&-:\g(z)\left(\hf\g(z)c(z)+C(z)\right)\beta(z):-:c(z)C(z)b(z):\nn
 &+&\kvt\left(\hf\g(z)c(z)+C(z)\right)\pa\var_1(z)-\kvt\left(\hf\g(z)c(z)-
  C(z)\right)\pa\var_2(z)\nn
 &+&\hf(k-1)\pa\g(z)c(z)-\hf(k+1)\pa c(z)\g(z)+k\pa C(z)
\label{Wakbos}
\eea
where we have introduced the simplifying notation
\bea
 \beta(z)&=&\beta_{\al_1}(z)\spa\g(z)=\g^{\al_1}(z)\nn
 b(z)&=&b_{\ald_2}(z)\hspace{.1cm}\spa c(z)=c^{\ald_2}(z)\nn
 B(z)&=&b_\ald(z)\hspace{.2cm}\spa C(z)=c^\ald(z) 
\eea
It is straightforward to verify that this is a free field realization
of the affine current superalgebra $A(1,0)^{(1)}$ with Cartan-Killing form 
and structure coefficients given by (\ref{A10G}) and (\ref{A10f}).
The Sugawara energy-momentum tensor is
\bea
 T(z)&=&:\pa\g(z)\beta(z):+:\pa c(z)b(z):+\pa C(z)B(z):\nn
 &+&\hf:\pa\var(z)\cdot\var(z):+\frac{1}{\kvt}\ald_2\cdot\pa^2\var(z)
\eea 
Owing to the equal numbers of bosonic and fermionic generators, the
central charge of the Sugawara tensor vanishes.

The screening currents of the first kind are easily found to be
\bea
 s_{\al_1}(z)&=&-\left(\beta(z)+\hf c(z)B(z)\right):e^{-\var_1(z)/\kvt}:\nn
 s_{\ald_2}(z)&=&-\left(b(z)-\hf\g(z)B(z)\right):e^{-\var_2(z)/\kvt}:
\eea
and it may be checked that indeed they have the required properties.
These particular expressions do not seem to have appeared in the literature
before.

Finally, the primary field of weight $\La$ is found to
be
\bea
 &&\phi_\La(w,x,\theta)\nn
 &=&\left[1+x^{\al_1}\g(w)+\left(\hf x^{\al_1}\theta^{\ald_2}+\theta^\ald
  \right)\left(\hf\g(w)c(w)+C(w)\right)\right]^{\La_1}\nn
 &\cdot&\left[1+\theta^{\ald_2}c(w)+\left(\hf x^{\al_1}
  \theta^{\ald_2}-\theta^\ald
  \right)\left(\hf \g(w)c(w)-C(w)\right)-2\theta^{\ald_2}\theta^\ald
  c(w)C(w)\right]^{\La_2}\nn
 &\cdot&:e^{\frac{1}{\kvt}\La\cdot\var(w)}:
\label{primdist}
\eea
This expression follows from the observation that the basis vectors
in $M_{\La^1}(A(1,0))$ and $M_{\La^2}(A(1,0))$ are proportional to
\ben
 \ket{\La^1}\ ,\ F_{\al_1}\ket{\La^1}\ ,\ f_\ald\ket{\La^1}=f_{\ald_2}
  F_{\al_1}\ket{\La^1}
\een
and
\ben
 \ket{\La^2}\ ,\ f_{\ald_2}\ket{\La^2}\ ,\ f_\ald\ket{\La^2}=-F_{\al_1}
  f_{\ald_2}\ket{\La^2} \ , \ f_{\ald_2}f_\ald\ket{\La^2}=-f_\ald 
  f_{\ald_2}\ket{\La^2}
\een
The explicit expression for the primary field with arbitrary weight
in (\ref{primdist}) is a new result illustrating the general construction
(\ref{primgen}) in Section 6.

Let us now turn to the alternative choice of a set of simple roots where
both simple roots are odd, $\ald_1$ and $\ald_2$. The remaining non-simple
root $\al=\ald_1+\ald_2$ is even. This alternative (purely odd) set of simple
roots, $\ald_1^{(2)}$ and $\ald_2^{(2)}$, is obtained from the (distinguished)
one used above, $\al_1^{(1)}$ and $\ald_2^{(1)}$, by Weyl reflections (see e.g.
\cite{FSS}) associated to
the odd root $\ald_2^{(1)}$. In particular, we find $\ald_1^{(2)}=\al_1^{(1)}
+\ald_2^{(1)}=\ald^{(1)}$ and $\ald_2^{(2)}=-\ald_2^{(1)}$ such that
$\al^{(2)}=\al_1^{(1)}$, and the Weyl vector becomes $\rho^{(2)}=0$. 
This implies that
\bea
 H_1^{(2)}&=&H_1^{(1)}+H_2^{(1)}\spa H_2^{(2)}=-H_2^{(1)}\nn
 e_{\ald_1}^{(2)}&=&e_{\ald}^{(1)}\hspace{1.5cm}
  \spa f_{\ald_1}^{(2)}=f_\ald^{(1)}\nn
 e_{\ald_2}^{(2)}&=&f_{\ald_2}^{(1)}\hspace{1.4cm}
  \spa f_{\ald_2}^{(2)}=-e_{\ald_2}^{(1)}\nn
 E_\al^{(2)}&=&E_{\al_1}^{(1)}\hspace{1.4cm}\spa F_\al^{(2)}=F_{\al_1}^{(1)}
\label{osc2}
\eea
Using this correspondence between the Lie superalgebra generators, one may
work out the structure coefficients
\bea
 {f_{\ald_1,-\ald_1}}^1=1&\spa&{f_{\ald_2,-\ald_2}}^2=1\nn
 \al(H_1)={f_{1,\al}}^\al=1&\spa&\al(H_2)={f_{2,\al}}^\al=1\nn
 {f_{\al,-\al}}^1=1&\spa&{f_{\al,-\al}}^2=1\nn
 {f_{\pm\ald_1,\mp\al}}^{\mp\ald_2}=1\spa&{f_{\pm\ald_1,\pm\ald_2}}^{\pm\al}
  =\pm1&\spa{f_{\pm\ald_2,\mp\al}}^{\mp\ald_1}=1
\eea
and in particular the Cartan-Killing form
\bea
 G_{11}=0\spa G_{12}&=&G_{21}=1\spa G_{22}=0\nn
 \kappa_{\ald_1,-\ald_1}&=&\kappa_{\ald_2,-\ald_2}=
  \kappa_{\al,-\al}=1
\eea
such that $A_{ij}=G_{ij}$.
Based on these parameters the alternative (fermionic)
free field realization becomes
\bea
 H_1(z)&=&-:c'(z)b'(z):-:\g(z)\beta(z):+\kvt\pa\var_1(z)\nn
 H_2(z)&=&-:c(z)b(z):-:\g(z)\beta(z):+\kvt\pa\var_2(z)\nn
 e_{\ald_1}(z)&=&b(z)+\hf c'(z)\beta(z)\nn
 f_{\ald_1}(z)&=&-:\left(\hf c(z)c'(z)+\g(z)\right)b'(z):
  -\hf c(z):\g(z)\beta(z):\nn
 &+&\kvt c(z)\pa\var_1(z)+(k+1/2)\pa c(z)\nn
 e_{\ald_2}(z)&=&b'(z)+\hf c(z)\beta(z)\nn
 f_{\ald_2}(z)&=&:\left(\hf c(z)c'(z)-\g(z)\right)b(z):-\hf c'(z):\g(z)
  \beta(z):\nn
 &+&\kvt c'(z)\pa\var_2(z)+(k+1/2)\pa c'(z)\nn
 E_\al(z)&=&\beta(z)\nn
 F_\al(z)&=&-:c(z)\g(z)b(z):-:c'(z)\g(z) b'(z):-:\g^2(z)\beta(z):\nn
 &-&\kvt\left(\hf c(z)c'(z)-\g(z)\right)\pa\var_1(z)
 +\kvt\left(\hf c(z)c'(z)+\g(z)\right)\pa\var_2(z)\nn
 &-&\hf(k+1)\pa c(z)c'(z)-\hf(k+1)\pa c'(z)c(z)+k\pa\g(z)
\label{Wakferm}
\eea
where we have introduced the shorthand notation
\bea
 b(z)&=&b_{\ald_1}(z)\spa c(z)=c^{\ald_1}(z)\nn
 b'(z)&=&b_{\ald_2}(z)\spa c'(z)=c^{\ald_2}(z)\nn
 \beta(z)&=&\beta_\al(z)\hspace{.1cm}\spa\g(z)=\g^\al(z)
\eea
Note that this Wakimoto realization is invariant under the interchanging
of $i=1$ and $i=2$.
The associated Sugawara energy-momentum tensor is
\ben
 T(z)=:\pa c(z)b(z):+:\pa c'(z)b'(z):+:\pa\g(z)\beta(z):+\hf:\pa\var(z)\cdot
  \pa\var(z):
\een
The screening currents of the first kind are
\bea
 s_{\ald_1}(z)&=&-\left(b(z)-\hf c'(z)\beta(z)\right):e^{-\var_1(z)/\kvt}:\nn
 s_{\ald_2}(z)&=&-\left(b'(z)-\hf c(z)\beta(z)\right):e^{-\var_2(z)/\kvt}:
\label{screenferm}
\eea
Similar realizations ((\ref{Wakferm}) and (\ref{screenferm})) in the 
fermionic basis of simple roots have also been obtained by Ito \cite{Ito3}.
More recently, in \cite{BKT} the relation is discussed between the Wakimoto
free field realizations (\ref{Wakbos}) and (\ref{Wakferm}) 
of the affine currents based on the two inequivalent choices of simple
roots.

Finally, the primary field of weight $\La$ becomes
\bea
 \phi_\La(w,x,\theta)&=&\left[1+\theta^{\ald_1}c(w)+\left(\hf\theta^{\ald_1}
  \theta^{\ald_2}-x^\al\right)\left(\hf c(w)c'(w)-\g(w)\right)
  \right]^{\La_1}\nn
 &\cdot&\left[1+\theta^{\ald_2}c'(w)+\left(\hf\theta^{\ald_1}\theta^{\ald_2}
  +x^\al\right)\left(\hf c(w)c'(w)+\g(w)\right)\right]^{\La_2}\nn
 &\cdot&:e^{\frac{1}{\kvt}\La\cdot\var(w)}:
\label{fermprim1}
\eea
Again, such an explicit result seems not to be found in the literature.
In this case the basis vectors are proportional to
\ben
 \ket{\La^1}\ ,\ f_{\ald_1}\ket{\La^1}\ ,\ F_\al\ket{\La^1}=-f_{\ald_2}
  f_{\ald_1}\ket{\La^1}
\een
and
\ben
 \ket{\La^2}\ ,\ f_{\ald_2}\ket{\La^2}\ ,\ F_\al\ket{\La^2}=-f_{\ald_1}
  f_{\ald_2}\ket{\La^2}
\een

Inspired by the oscillator representation (\ref{osc}) and (\ref{osc2})
we may choose the fundamental labels differently, namely as
\ben
 \La^1(H_1)=1\ ,\ \La^1(H_2)=0\ ,\ \La^2(H_1)=2\ ,\ \La^2(H_2)=-1
\een
as opposed to the diagonal choice above: $\La^k(H_i)=\delta^k_i$.
Based on this new set $\left\{\La^k\right\}$ we find the primary field
\bea
  \phi_\La(w,x,\theta)&=&\left[1+\theta^{\ald_1}c(w)+\left(\hf\theta^{\ald_1}
  \theta^{\ald_2}-x^\al\right)\left(\hf c(w)c'(w)-\g(w)\right)
  \right]^{\La_1+2\La_2}\nn
 &\cdot&\left[1+2\theta^{\ald_1}c(w)-\theta^{\ald_2}c'(w)
  +\left(\frac{3}{2}\theta^{\ald_1}\theta^{\ald_2}
  -x^\al\right)\left(\frac{3}{2}c(w)c'(w)-\g(w)\right)\right]^{-\La_2}\nn
 &\cdot&:e^{\frac{1}{\kvt}\La\cdot\var(w)}:
\label{fermprim2}
\eea
Again, such an explicit result seems not to be found in the literature.
In this case the basis vectors are proportional to
\ben
 \ket{\La^1}\ ,\ f_{\ald_1}\ket{\La^1}\ ,\ F_\al\ket{\La^1}
\een
and
\ben
 \ket{\La^2}\ ,\ f_{\ald_1}\ket{\La^2}\ ,\
  f_{\ald_2}\ket{\La^2}\ ,\ F_\al\ket{\La^2}
\een
As a consistency check it is easily verified that for given (basis
independent) labels $\La_1$ and $\La_2$ of $\La$, the two primary fields
(\ref{fermprim1}) and (\ref{fermprim2}) are identical. This illustrates the 
freedom in choosing a convenient basis $\left\{\La^k\right\}$ in weight space.

In \cite{SC} differential operator realizations and free field 
realizations are discussed in the case of the Lie supergroup $OSp(2|2)$. 
We note that the realizations obtained there are different from ours (even for
the Sugawara tensor) and are based on a different set of defining
commutator relations for the Lie superalgebra $osp(2|2)$. We don't know
the translations between them.

\section{Conclusions}

In this paper we have provided in particular 4 new results. First, we have
derived general differential operator realizations of Lie superalgebras.
Second, we have {\em quantized} these {\em classical} realizations and
thereby obtained general free field realizations of affine current
superalgebras. In this process the non-trivial part was to take care of
multiple contractions by adding anomalous terms to the lowering operators.
Third, we have worked out general expressions for screening currents of
the first kind and presented proofs of their properties. Fourth,
we have provided explicit generating function primary fields for arbitrary
representations, based on super-triangular coordinates.
Finally, we have compared the results with the literature
and found them to be in accordance. 

The results allow setting up integral representations for correlators
of primary fields corresponding to integrable representations. When screening
currents of the second kind have been worked out, we then have sufficient
ingredients for setting up integral representations for correlators
in the case of degenerate representations. This would be of interest
e.g. in the $G/G$ approach to non-critical strings. In the case of 
$OSp(1|2)$ the screening current of the second kind is known \cite{ERdeS}.
We intend to come back elsewhere
with a discussion of screening currents of the second kind in more
general situations \cite{JR3} and of
correlators for degenerate representations of $OSp(1|2)$ \cite{PRSY}.

Very recently there has been rapidly increasing interest in realizations 
of $q$-deformed Lie (super-)algebras. However, most results only pertain
to specific examples (see e.g. \cite{Kim,SC} and references therein). 
It would be interesting to try to develop
a general scheme for obtaining such realizations.\\[.8 cm] 
{\bf Acknowledgement}\\[.2cm]
The author wants to express his sincere gratitude towards J.L. Petersen
and M. Yu for the collaboration on the recent common
paper \cite{PRY3} (and also on \cite{PRY1,PRY2}) of
which the present work is a generalization.
He also thanks J.L. Petersen and H.-T. Sato for fruitful discussions
on $osp(1|2)$ in the early stages of this work, and L. Feh\'er for pointing
out some references.

\appendix
\section{Polynomial Identities}

This appendix is devoted to listing several polynomial 
identities following from the realizations obtained in Section 3 and 
Section 4, and used in Section 5 and Section 6. 
The terminology of {\em classical} and {\em quantum} polynomial
identities originates in our viewpoint of affine current superalgebras
being quantizations of the corresponding (classical) Lie superalgebras. 
Thus, identities obtained from comparison of the differential operator
algebra with the Lie superalgebra (anti-)commutator relations, are denoted
classical polynomial identities. 
Similarly, the identities obtained from comparing the free field
realization with the defining OPE of the affine current superalgebra, are
denoted quantum polynomial identities. 

In this appendix we leave out the arguments of the polynomials.
All root indices represent positive roots, while $a$ ($\dot{a}$) represents
any even root or Cartan index (any odd root).
 
\subsection{Classical Polynomial Identities}

{}From the differential operator realization of the Lie superalgebra we
derive the following set of classical polynomial identities
\bea
 \pm\aldd(H_i)V_{\pm\aldd}^\sigmadd&=&V_i^\gdd\pa_\gdd V_{\pm\aldd}^\sigmadd-
  V_{\pm\aldd}^\gdd\pa_\gdd V_i^\sigmadd\nn
 -\aldd(H_i)P_{-\aldd}^j&=&V_i^\gdd\pa_\gdd P_{-\aldd}^j\nn
 {f_{\al,\beta}}^\g V_\g^\sigmadd&=&V_\al^\gdd\pa_\gdd V_\beta^\sigmadd-
  V_\beta^\gdd\pa_\gdd V_\al^\sigmadd\nn
  {f_{\al,-\beta}}^aV_a^\sigmadd&=&V_\al^\gdd\pa_\gdd V_{-\beta}^\sigmadd-
  V_{-\beta}^\gdd\pa_\gdd V_\al^\sigmadd\nn
 {f_{\aldd,-\aldd}}^j&=&V_\aldd^\gdd\pa_\gdd P_{-\aldd}^j\nn
 {f_{\al,-\beta}}^{-\g}P_{-\g}^j&=&V_\al^\gdd\pa_\gdd P_{-\beta}^j
  \hspace{1.5cm}\spa\mbox{for}\ \ \beta-\al\in\D_+^0\nn
 {f_{\al,\ald}}^\betad V_\betad^\sigmadd&=&V_\al^\gdd\pa_\gdd V_\ald^\sigmadd
  -V_\ald^\gdd\pa_\gdd V_\al^\sigmadd\nn
 {f_{\al,-\ald}}^{\dot{a}}V_{\dot{a}}^\sigmadd&=&V_\al^\gdd\pa_\gdd 
  V_{-\ald}^\sigmadd-V_{-\ald}^\gdd\pa_\gdd V_\al^\sigmadd\nn
 {f_{\al,-\ald}}^{-\betad}P_{-\betad}^j&=&V_\al^\gdd\pa_\gdd P_{-\ald}^j
  \hspace{1.5cm}\spa\mbox{for}\ \ \ald-\al\in\D_+^1\nn
 {f_{-\al,-\beta}}^{-\g}V_{-\g}^\sigmadd&=&V_{-\al}^\gdd\pa_\gdd
  V_{-\beta}^\sigmadd-V_{-\beta}^\gdd\pa_\gdd V_{-\al}^\sigmadd\nn
 {f_{-\al,-\beta}}^{-\g}P_{-\g}^j&=&V_{-\al}^\gdd\pa_\gdd
  P_{-\beta}^j-V_{-\beta}^\gdd\pa_\gdd P_{-\al}^j\nn
 {f_{\ald,-\al}}^{\dot{a}}V_{\dot{a}}^\sigmadd&=&
  V_\ald^\gdd\pa_\gdd V_{-\al}^\sigmadd-V_{-\al}^\gdd\pa_\gdd
  V_\ald^\sigmadd\nn
 {f_{\ald,-\al}}^{-\betad}P_{-\betad}^j&=&V_\ald^\gdd\pa_\gdd
  P_{-\al}^j
 \hspace{1.5cm}\spa\mbox{for}\ \ \al-\ald\in\D_+^1\nn
 {f_{\ald,\betad}}^\g V_\g^\sigmadd&=&V_\ald^\gdd\pa_\gdd V_\betad^\sigmadd
  +V_\betad^\gdd\pa_\gdd V_\ald^\sigmadd\nn
 {f_{\ald,-\betad}}^a V_a^\sigmadd&=&V_\ald^\gdd\pa_\gdd V_{-\betad}^\sigmadd
  +V_{-\betad}^\gdd\pa_\gdd V_\ald^\sigmadd\nn
 {f_{\ald,-\betad}}^{-\g}P_{-\g}^j&=&V_\ald^\gdd\pa_\gdd P_{-\betad}^j
  \hspace{1.5cm}\spa\mbox{for}\ \ \betad-\ald\in\D_+^0\nn
 {f_{-\al,-\ald}}^{-\betad}V_{-\betad}^\sigmadd&=&V_{-\al}^\gdd\pa_\gdd
  V_{-\ald}^\sigmadd-V_{-\ald}^\gdd\pa_\gdd V_{-\al}^\sigmadd\nn
 {f_{-\al,-\ald}}^{-\betad}P_{-\betad}^j&=&V_{-\al}^\gdd\pa_\gdd
  P_{-\ald}^j-V_{-\ald}^\gdd\pa_\gdd P_{-\al}^j\nn
 {f_{-\ald,-\betad}}^{-\g}V_{-\g}^\sigmadd&=&V_{-\ald}^\gdd\pa_\gdd
  V_{-\betad}^\sigmadd+V_{-\betad}^\gdd\pa_\gdd V_{-\ald}^\sigmadd\nn
 {f_{-\ald,-\betad}}^{-\g}P_{-\g}^j&=&V_{-\ald}^\gdd\pa_\gdd
  P_{-\betad}^j+V_{-\betad}^\gdd\pa_\gdd P_{-\ald}^j
\label{class}
\eea
Of course, this set of relations can be written in a more compact way using
obvious properties such as $\al-\ald\neq0$. In Section 5 and Section 6
the following (``compact'') recursion relations prove themselves useful:
\bea
 {f_{\aldd,\betadd}}^\gdd V_\gdd^\sigmadd&=&V_\aldd^\gdd\pa_\gdd
  V_\betadd^\sigmadd-(-1)^{p(\aldd)p(\betadd)}V_\betadd^\gdd\pa_\gdd
  V_\aldd^\sigmadd\nn
 {f_{\aldd,-\betadd}}^{\pm\gdd}V_{\pm\gdd}^\sigmadd&=&V_\aldd^\gdd\pa_\gdd
  V_{-\betadd}^\sigmadd-(-1)^{p(\aldd)p(\betadd)}V_{-\betadd}^\gdd\pa_\gdd
  V_\aldd^\sigmadd\nn
 {f_{-\aldd,-\betadd}}^{-\gdd}V_{-\gdd}^\sigmadd&=&V_{-\aldd}^\gdd\pa_\gdd
  V_{-\betadd}^\sigmadd-(-1)^{p(\aldd)p(\betadd)}
  V_{-\betadd}^\gdd\pa_\gdd V_{-\aldd}^\sigmadd\nn
 {f_{\aldd,-\betadd}}^{-\gdd}P_{-\gdd}^j&=&V_\aldd^\gdd\pa_\gdd P_{-\betadd}^j
  \nn
 {f_{-\aldd,-\betadd}}^{-\gdd}P_{-\gdd}^j&=&V_{-\aldd}^\gdd\pa_\gdd
  P_{-\betadd}^j-(-1)^{p(\aldd)p(\betadd)}
  V_{-\betadd}^\gdd\pa_\gdd P_{-\aldd}^j
\label{rec1}
\eea
Similarly, from the differential screening operator commutation relations
(\ref{Scomm}) we find
\bea
 V_\al^\gdd\pa_\gdd S_\betadd^\sigmadd-S_\betadd^\gdd\pa_\gdd V_\al^\sigmadd
  &=&0\nn
 V_i^\gdd\pa_\gdd S_\betadd^\sigmadd-S_\betadd^\gdd\pa_\gdd V_i^\sigmadd
  &=&\betadd(H_i)S_\betadd^\sigmadd\nn
 V_{-\aldd}^\gdd\pa_\gdd S_\beta^\sigmadd-S_\beta^\gdd\pa_\gdd 
  V_{-\aldd}^\sigmadd&=&\beta(H_j)P_{-\aldd}^j S_\beta^\sigmadd
  -{f_{\beta,-\g}}^\mu Q_{-\aldd}^{-\g}S_\mu^\sigmadd-
  {f_{\beta,-\gd}}^\mud Q_{-\aldd}^{-\gd}S_\mud^\sigmadd\nn
 S_\beta^\gdd\pa_\gdd P_{-\aldd}^j&=&-{f_{\beta,-\g}}^j Q_{-\aldd}^{-\g}\nn
 V_{-\aldd}^\gdd\pa_\gdd S_\betad^\sigmadd
  -(-1)^{p(\aldd)}S_\betad^\gdd\pa_\gdd 
  V_{-\aldd}^\sigmadd&=&\betad(H_j)P_{-\aldd}^j S_\betad^\sigmadd
  -{f_{\betad,-\gd}}^\mu Q_{-\aldd}^{-\gd}S_\mu^\sigmadd-
  {f_{\betad,-\g}}^\mud Q_{-\aldd}^{-\g}S_\mud^\sigmadd\nn
 S_\betad^\gdd\pa_\gdd P_{-\aldd}^j&=&(-1)^{p(\aldd)}
  {f_{\betad,-\gd}}^jQ_{-\aldd}^{-\gd}\nn
 V_\ald^\gdd\pa_\gdd S_\betadd^\sigmadd-(-1)^{p(\betadd)}
  S_\betadd^\gdd\pa_\gdd V_\ald^\sigmadd&=&0\nn
 S_\al^\gdd\pa_\gdd S_\beta^\sigmadd-S_\beta^\gdd\pa_\gdd S_\al^\sigmadd&=&
  {f_{\al,\beta}}^\g S_\g^\sigmadd\nn
 S_\al^\gdd\pa_\gdd S_\betad^\sigmadd-S_\betad^\gdd\pa_\gdd S_\al^\sigmadd&=&
  {f_{\al,\betad}}^\gd S_\gd^\sigmadd\nn
 S_\ald^\gdd\pa_\gdd S_\betad^\sigmadd+
  S_\betad^\gdd\pa_\gdd S_\ald^\sigmadd&=&
  {f_{\ald,\betad}}^\g S_\g^\sigmadd
\label{Sclass}
\eea
In the proof in Section 5 the following are useful ``compactified'' relations
\bea
 V_{-\aldd}^\gdd\pa_\gdd S_{\aldd_j}^\sigmadd-(-1)^{p(\aldd_j)p(\aldd)}
  S_{\aldd_j}^\gdd\pa_\gdd V_{-\aldd}^\sigmadd&=&A_{ij}P_{-\aldd}^i
  S_{\aldd_j}^\sigmadd\nn
 S_{\aldd_j}^\gdd\pa_\gdd P_{-\aldd}^i&=&-(-1)^{p(\aldd_j)(1-p(\aldd))}
  \delta_j^iQ_{-\aldd}^{-\aldd_j}
\label{Sclass2}
\eea

\subsection{Quantum Polynomial Identities}

{}From the free field realization of the affine current superalgebra (and
using the classical polynomial identities (\ref{class})) we derive
the following set of quantum polynomial identities
\bea
 \betadd(H_i)F_{\betadd\gdd}&=&\gdd(H_i)F_{\betadd\gdd}-
  V_i^\sigmadd\pa_\sigmadd F_{\betadd\gdd}\nn
 0&=&tG_{ij}P_{-\betadd}^j+V_i^\gdd F_{\betadd\gdd}+(-1)^{p(\betadd)}
  \pa_\sigmad V_i^\gdd\pa_\gdd V_{-\betadd}^\sigmad-\pa_\sigma
  V_i^\gdd\pa_\gdd V_{-\betadd}^\sigma\nn
 k\kappa_{\al,-\beta}&=&-\pa_\sigma V_\al^\gdd\pa_\gdd V_{-\beta}^\sigma
  +\pa_\sigmad V_\al^\gdd\pa_\gdd V_{-\beta}^\sigmad
  +V_\al^\gdd F_{\beta\gdd}\nn
 k\kappa_{\al,-\betad}&=&-\pa_\sigmadd V_\al^\gdd
  \pa_\gdd V_{-\betad}^\sigmadd+V_\al^\gdd f_{\betad\gdd}\nn
  k\kappa_{\ald,-\beta}&=&-\pa_\sigmadd V_\ald^\gdd\pa_\gdd V_{-\beta}^\sigmadd
  +V_\ald^\gdd F_{\beta\gdd}\nn
 k\kappa_{\ald,-\betad}&=&-\pa_\sigma V_\ald^\gdd\pa_\gdd V_{-\betad}^\sigma
  +\pa_\sigmad V_\ald^\gdd\pa_\gdd V_{-\betad}^\sigmad
  +V_\ald^\gdd f_{\betad\gdd}\nn
 {f_{\aldd,-\betadd}}^{-\gdd}F_{\gdd\sigmadd}&=&(-1)^{p(\aldd)p(\sigmadd)}
  V_\aldd^\gdd\pa_\gdd F_{\betadd\sigmadd}+\pa_\sigmadd V_\aldd^\gdd
  F_{\betadd\gdd}\nn
 &-&\pa_\mu\pa_\sigmadd V_\aldd^\gdd\pa_\gdd V_{-\betadd}^\mu
  +(-1)^{p(\aldd)+p(\betadd)+p(\sigmadd)}\pa_\mud\pa_\sigmadd
  V_\aldd^\gdd\pa_\gdd V_{-\betadd}^\mud\nn
 0&=&-\pa_\sigma V_{-\al}^\gdd\pa_\gdd V_{-\beta}^\sigma+\pa_\sigmad
  V_{-\al}^\gdd\pa_\gdd V_{-\beta}^\sigmad+tG_{ij}P_{-\al}^iP_{-\beta}^j
  +V_{-\al}^\gdd F_{\beta\gdd}+V_{-\beta}^\gdd F_{\al\gdd}\nn
 {f_{-\al,-\beta}}^{-\g}F_{\g\sigmadd}&=&-\pa_\mu\pa_\sigmadd V_{-\al}^\gdd
  \pa_\gdd V_{-\beta}^\mu+(-1)^{p(\sigmadd)}\pa_\mud\pa_\sigmadd V_{-\al}^\gdd
  \pa_\gdd V_{-\beta}^\mud+tG_{ij}\pa_\sigmadd P_{-\al}^iP_{-\beta}^j\nn
 &+&V_{-\al}^\gdd\pa_\gdd F_{\beta\sigmadd}-V_{-\beta}^\gdd\pa_\gdd
  F_{\al\sigmadd}+\pa_\sigmadd V_{-\al}^\gdd F_{\beta\gdd}
  +V_{-\beta}^\g\pa_\sigmadd F_{\al\g}+(-1)^{p(\sigmadd)}V_{-\beta}^\gd
  \pa_\sigmadd F_{\al\gd}\nn
 0&=&-\pa_\sigmadd V_{-\al}^\gdd\pa_\gdd V_{-\betad}^\sigmadd+tG_{ij}
  P_{-\al}^iP_{-\betad}^j+V_{-\al}^\gdd f_{\betad\gdd}+V_{-\betad}^\gdd
  F_{\al\gdd}\nn
 {f_{-\al,-\betad}}^{-\gd}f_{\gd\sigmadd}&=&-\pa_\mu\pa_\sigmadd
  V_{-\al}^\gdd\pa_\gdd V_{-\betad}^\mu-(-1)^{p(\sigmadd)}\pa_\mud\pa_\sigmadd
  V_{-\al}^\gdd\pa_\gdd V_{-\betad}^\mud+tG_{ij}\pa_\sigmadd P_{-\al}^i
  P_{-\betad}^j\nn
 &+&V_{-\al}^\gdd\pa_\gdd f_{\betad\sigmadd}-(-1)^{p(\sigmadd)}V_{-\betad}^\gdd
  \pa_\gdd F_{\al\sigmadd}+\pa_\sigmadd V_{-\al}^\gdd f_{\betad\gdd}\nn
 &+&(-1)^{p(\sigmadd)}V_{-\betad}^\g\pa_\sigmadd F_{\al\g}+V_{-\betad}^\gd
  \pa_\sigmadd F_{\al\gd}\nn
 0&=&-\pa_\sigmadd V_{-\ald}^\gdd\pa_\gdd V_{-\beta}^\sigmadd +tG_{ij}
  P_{-\ald}^iP_{-\beta}^j+V_{-\ald}^\gdd F_{\beta\gdd}+V_{-\beta}^\gdd
  f_{\ald\gdd}\nn
 {f_{-\ald,-\beta}}^{-\gd}f_{\gd\sigmadd}&=&-\pa_\mu\pa_\sigmadd
  V_{-\ald}^\gdd\pa_\gdd V_{-\beta}^\mu-(-1)^{p(\sigmadd)}\pa_\mud
  \pa_\sigmadd V_{-\ald}^\gdd\pa_\gdd V_{-\beta}^\mud+tG_{ij}\pa_\sigmadd
  P_{-\ald}^iP_{-\beta}^j\nn
 &+&(-1)^{p(\sigmadd)}V_{-\ald}^\gdd\pa_\gdd F_{\beta\sigmadd}
  -V_{-\beta}^\gdd\pa_\gdd f_{\ald\sigmadd}+\pa_\sigmadd V_{-\ald}^\gdd
  F_{\beta\gdd}\nn
 &+&V_{-\beta}^\g\pa_\sigmadd f_{\ald\g}+(-1)^{p(\sigmadd)}V_{-\beta}^\gd
  \pa_\sigmadd f_{\ald\gd}\nn
 0&=&-\pa_\sigma V_{-\ald}^\gdd\pa_\gdd V_{-\betad}^\sigma+\pa_\sigmad
  V_{-\ald}^\gdd\pa_\gdd V_{-\betad}^\sigmad+tG_{ij}P_{-\ald}^iP_{-\betad}^j
  +V_{-\ald}^\gdd f_{\betad\gdd}-V_{-\betad}^\gdd f_{\ald\gdd}\nn
 {f_{-\ald,-\betad}}^{-\g}F_{\g\sigmadd}&=&-\pa_\mu\pa_\sigmadd
  V_{-\ald}^\gdd\pa_\gdd V_{-\betad}^\mu+(-1)^{p(\sigmadd)}\pa_\mud\pa_\sigmadd
  V_{-\ald}^\gdd\pa_\gdd V_{-\betad}^\mud+tG_{ij}\pa_\sigmadd P_{-\ald}^i
  P_{-\betad}^j\nn
 &+&(-1)^{p(\sigmadd)}\left(V_{-\ald}^\gdd\pa_\gdd f_{\betad\sigmadd}
  +V_{-\betad}^\gdd\pa_\gdd f_{\ald\sigmadd}\right)
  +\pa_\sigmadd V_{-\ald}^\gdd f_{\betad\gdd}\nn
 &-&(-1)^{p(\sigmadd)}V_{-\betad}^\g\pa_\sigmadd f_{\ald\g}
  -V_{-\betad}^\gd\pa_\sigmadd f_{\ald\gd}\nn
 0&=&\pa_\g V_{-\aldd}^\g-(-1)^{p(\aldd)}
  \pa_\gd V_{-\aldd}^\gd+2\rho(H_j)P_{-\aldd}^j 
\label{quant}
\eea
Some of the terms obviously vanish since
$\kappa_{\aldd,-\betadd}=0$ unless $\aldd=\betadd$. We use the common
notation $F_{\aldd\betadd}$ for the anomalous terms $F_{\al\betadd}$
and $f_{\ald\betadd}$ when $\aldd$ is an arbitrary positive root.
We may re-express some of these identities as the following (``compact'')
recursion relation 
\bea
 {f_{-\aldd,-\betadd}}^{-\gdd}F_{\gdd\sigmadd}&=&-\pa_\mu\pa_\sigmadd
  V_{-\aldd}^\gdd\pa_\gdd V_{-\betadd}^\mu+(-1)^{p(\aldd)+p(\betadd)+
  p(\sigmadd)}\pa_\mud\pa_\sigmadd V_{-\aldd}^\gdd\pa_\gdd V_{-\betadd}^\mud
  +tG_{ij}\pa_\sigmadd P_{-\aldd}^iP_{-\betadd}^j\nn
 &+&(-1)^{p(\aldd)p(\sigmadd)}V_{-\aldd}^\gdd\pa_\gdd F_{\betadd\sigmadd}
  -(-1)^{p(\betadd)(p(\aldd)-p(\sigmadd))}V_{-\betadd}^\gdd\pa_\gdd
  F_{\aldd\sigmadd}+\pa_\sigmadd V_{-\aldd}^\gdd F_{\betadd\gdd}\nn
 &+&(-1)^{p(\betadd)(p(\aldd)-p(\sigmadd))}V_{-\betadd}^\g\pa_\sigmadd
  F_{\aldd\g}+(-1)^{p(\aldd)p(\betadd)+(1-p(\betadd))p(\sigmadd)}
  V_{-\betadd}^\gd\pa_\sigmadd F_{\aldd\gd}
\label{rec2}
\eea
and as
\bea
 0&=&-\pa_\mu V_{-\aldd}^\gdd\pa_\gdd V_{-\betadd}^\mu +(-1)^{p(\aldd)
  +p(\betadd)}\pa_\mud V_{-\aldd}^\gdd\pa_\gdd V_{-\betadd}^\mud
  +tG_{ij}P_{-\aldd}^iP_{-\betadd}^j\nn
 &+&V_{-\aldd}^\gdd
  F_{\betadd\gdd}+(-1)^{p(\aldd)p(\betadd)}V_{-\betadd}^\gdd F_{\aldd\gdd}
\label{quant2}
\eea
It is not difficult to show that (\ref{quant2}) follows from (\ref{rec2}).

\end{document}